\documentclass[twocolumn, twocolappendix]{aastex631}

\begin{document}

\title{Corrected SFD: A More Accurate Galactic Dust Map with Minimal Extragalactic Contamination}

\shorttitle{CSFD Dust Map}
\shortauthors{Chiang}

\email{ykchiang@asiaa.sinica.edu.tw}

\author[0000-0001-6320-261X]{Yi-Kuan Chiang}
\affiliation{Academia Sinica Institute of Astronomy and Astrophysics (ASIAA), No. 1, Section 4, Roosevelt Road, Taipei 10617, Taiwan}

\begin{abstract}
The widely used Milky Way dust reddening map, the \citet*[][SFD]{1998ApJ...500..525S} map, was found to contain extragalactic large-scale structure (LSS) imprints \citep*{2019ApJ...870..120C}. Such contamination is inherent in maps based on infrared emission, which pick up not only Galactic dust but also the cosmic infrared background (CIB). When SFD is used for extinction correction, overcorrection occurs in a spatially correlated and redshift-dependent manner, which could impact precision cosmology using galaxy clustering, lensing, and supernova Ia distances. Similarly, LSS imprints in other Galactic templates can affect intensity mapping and cosmic microwave background experiments. This paper presents a generic way to remove LSS traces in Galactic maps and applies it to SFD. First, we measure descriptive summary statistics of the CIB in SFD by cross-correlating the map with spectroscopic galaxies and quasars in SDSS tomographically as functions of redshift and angular scale. To reconstruct the LSS on the map level, however, additional information on the phases is needed. We build a large set of 180 overcomplete, full-sky basis template maps from the density fields of over 600 million galaxies in WISE and find a linear combination that reproduces all of the high-dimensional tomographic two-point statistics of the CIB in SFD. After subtracting this reconstructed LSS/CIB field, the end product is a full-sky Galactic dust reddening map that supersedes SFD, carrying all Galactic features therein, with maximally suppressed CIB. We release this new dust map dubbed CSFD---the Corrected SFD---at \url{https://idv.sinica.edu.tw/ykchiang/CSFD.html} and NASA's LAMBDA  archive. 
\end{abstract}

\section{Introduction} \label{sec:intro}

Observations of the Universe require seeing through our Milky Way galaxy, where extinction and reddening occur due to wavelength-dependent absorption and scattering by dust grains in the interstellar medium (ISM). As there is no single sight line in the sky without a detectable dust column, Galactic extinction correction is needed for all extragalactic photometry in the ultraviolet (UV), optical, and near-infrared (NIR).

So far, the most widely used Galactic dust reddening map has been the \cite*{1998ApJ...500..525S} map (hereafter SFD). It is primarily based on the nearly full-sky 100~$\mu$m intensity map taken by the Infrared Astronomical Satellite \citep[IRAS;][]{1984ApJ...278L...1N}. To probe the extinction, which scales with dust column density, a temperature correction was applied using the 100 and 240~$\mu$m maps from the Diffuse Infrared Background Experiment (DIRBE) on the Cosmic Background Explorer \citep{1992ApJ...397..420B}. The SFD map is provided in reddening, or color excess E($B-V$) scale, with a beam of $6.1'$ FHWM set by the reprocessed IRAS data. There is a small fraction (2\%) of the sky with no IRAS data, where SFD carries the DIRBE resolution at 1$^{\circ}$. To apply a wavelength-dependent extinction correction, one needs to further combine the color excess with an extinction curve. \cite{1998ApJ...500..525S} originally suggested using the \cite{1989ApJ...345..245C} law, while \cite{2011ApJ...737..103S} showed that the \cite{1999PASP..111...63F} law is more accurate and provided per-band coefficients derived accordingly.

The SFD dust map has been routinely used in extragalactic astronomy and cosmology. The Sloan Digital Sky Survey \citep[SDSS;][]{2000AJ....120.1579Y} uses SFD for Galactic extinction since the very first, early data release \citep{2002AJ....123..485S}. In probing the cosmic expansion, the construction of Hubble diagrams using supernovae Ia as standard candles requires the Ia's to be dereddened \citep[e.g., Pantheon;][]{2018ApJ...859..101S,2022ApJ...938..110B}. Looking forward, Dark Energy Spectroscopic Instrument \citep[DESI;][]{2016arXiv161100036D}, Prime Focus Spectrograph Subaru Strategic Program \citep{2014PASJ...66R...1T}, Rubin Observatory Legacy Survey of Space and Time \citep{2019ApJ...873..111I}, and other Stage IV cosmology experiments all plan to use SFD for extinction correction.

With intensive use by the community, the precision and accuracy of SFD need to be scrutinized. \cite{2011ApJ...737..103S} recalibrated the amplitude of the Galactic reddening using stellar spectra and found that the SFD color excess needs to be down-scaled by a factor of 0.86 (i.e., $14\%$). It has also been pointed out that the SFD map, constructed from far-infrared (FIR) emission, contains detectable traces of extragalactic large-scale structure (LSS) from the unresolved cosmic infrared background (CIB) originating from the same dust physics in external galaxies. \cite{2007PASJ...59..205Y} first showed that over a large patch of the high-latitude sky, the surface number density of SDSS galaxies increases with the E($B-V$) reported in SFD. This is the opposite of the physical extinction effect one would expect but is consistent with the SFD reddening field being biased in a spatial-dependent way due to the CIB contamination. This is further confirmed by \cite{2013PASJ...65...43K}, who stacked the SFD map at the locations of photometric galaxies in SDSS and found significant detections. 

Recently, \citet[][hereafter CM19]{2019ApJ...870..120C} performed tomographic diagnostics for 10 Galactic dust maps, including SFD. They characterized the CIB, or LSS contamination of other origins in these maps via cross-correlating with spectroscopic galaxies and quasars in SDSS and extracted the redshift-dependent statistics. They showed that on scales close to the beam, the amplitude of the CIB in SFD is the highest at low redshifts, reaching the level of over a percent of the values in the map. Meanwhile, the high-redshift tail is detectable up to $z\sim2$. 

Dust map systematics is concerning for a wide range of upcoming cosmology experiments aiming at precision constraints in, e.g., matter density, growth rates, dark energy, neutrino masses, and primordial non-Gaussianity. When SFD is used to derive absolute photometry for sources, due to the CIB residual, Galactic extinction would be overcorrected in a redshift- and spatial-dependent manner following the pattern of the cosmic web. These correlated overcorrections then bias estimators in cosmology either directly (e.g., standard candle magnitudes or lensing magnification factors) or indirectly via modulating the threshold fluxes for sample selection, thereby affecting all source-based analyses. These biases can then mimic the effect of cosmology in, e.g., supernovae Ia Hubble diagram, galaxy clustering, and gravitational lensing \citep[CM19;][]{2016ApJ...829...50A, 2020MNRAS.496.2262K}.

Interestingly, CM19 also showed that when extracting the CIB in SFD using different populations of galaxies as cross-correlation references, the cross-amplitudes scale with the bias factors of the galaxies but not their star-formation rates. This suggests that, at the resolution of SFD, the CIB is dominated by the so-called ``two-halo'' term, i.e., 100~$\rm \mu m$ light from star-forming galaxies clustered with the references but not the reference galaxies themselves. This implies that it is possible to clean the CIB in SFD using some generic LSS tracers in sky surveys in the optical or NIR without needing to resolve every galaxy in the FIR.  

This paper presents a tomographic reconstruction of the LSS, or the CIB field in SFD, and subtracts it to derive a new Galactic dust map dubbed CSFD---the corrected SFD. This CIB reconstruction is fully empirical, and no explicit modeling on galaxy--dark matter halo connection is needed. We provide a summary schematic in Figure~\ref{fig:Summary}-A. By construction, CSFD preserves all major properties of SFD but with minimal spatial systematics that correlates with the extragalactic LSS. We introduce our method in Section~\ref{sec:formalism} and apply it to separate SFD into the pure Galactic CSFD plus a reconstructed CIB map in Section~\ref{sec:comp_sep_SFD}. We validate these map products in Sections~\ref{sec:validation} before discussing and concluding our work in Sections~\ref{sec:discuss} and \ref {sec:conclusion}, respectively. Data preparation and a substantial amount of detail in optimizing the reconstruction are presented in the Appendix for interested readers. In setting the redshift- and spatial-binning scheme, for convenience, we adopt a $\rm \Lambda$ cold dark matter cosmology using parameters in \cite{2020A&A...641A...6P}: ($h$, $\Omega_{\rm c}h^2$, $\Omega_{\rm b}h^2$, $A_{\rm s}$, $n_{\rm s}$) = (0.6737, 0.1198, 0.02233, $2.097\times10^{-9}$, 0.9652), while we note that the end products, both the reconstructed LSS map and the CSFD dust map are entirely cosmology-independent.

\section{Formalism} \label{sec:formalism}

Here we introduce a generic formalism for Galactic-extragalactic component separation before applying it to the case of SFD cleaning in the next Section.

Let us consider a sky map of an arbitrary quantity $I$ as a function of angle $\phi$. Naturally, this map could contain everything along the line of sight, including the Galactic foreground, $I_{G}(\phi)$, the extragalactic LSS, $I_{LSS}(\phi)$, and potentially the early Universe background, $I_{BG}(\phi)$: 
\begin{eqnarray}
I(\phi) &=& I_{G}(\phi) + I_{LSS}(\phi) + I_{BG}(\phi) \nonumber \\
        &\approx& I_{G}(\phi) + I_{LSS}(\phi), 
\label{eq:1}
\end{eqnarray}
where $I_{BG}$ is often negligible except in the cosmic microwave background (CMB) experiments. We restrict our discussion to the regime where components add linearly,\footnote{Nonlineality can be introduced, e.g., in the UV and blue optical, where Milky Way dust contributes to $I_{G}$ via scatter light but also extincts the cosmic UV background \citep{2019ApJ...877..150C}.} as is the case for SFD.

\begin{figure*}[t!]
    \begin{center}
         \includegraphics[width=1\textwidth]{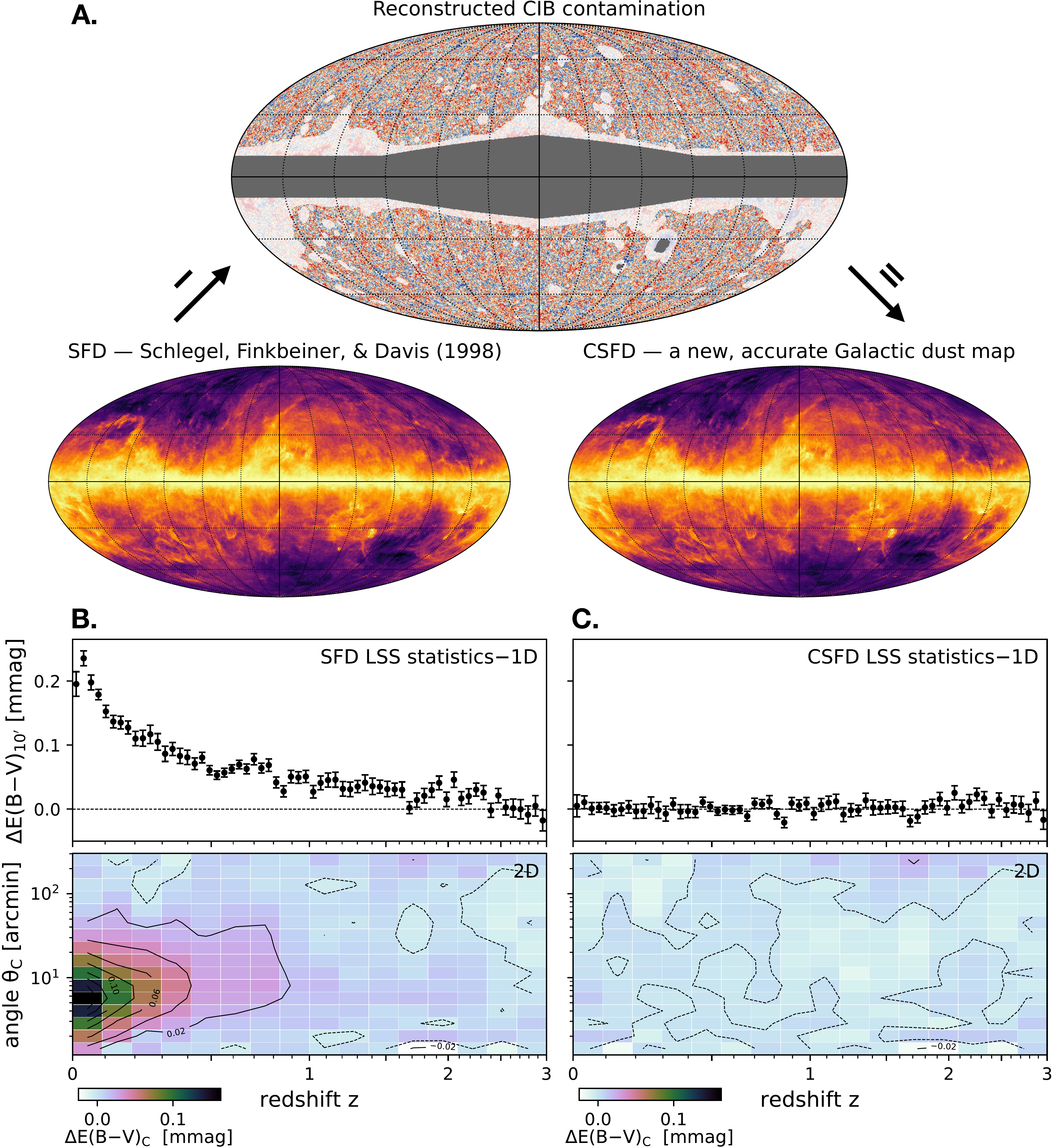}
    \end{center}
    \vspace{-1.5mm}
    \caption{\textbf{Panel A:} Summary of this work. We take the \citet[][SFD]{1998ApJ...500..525S} Galactic dust reddening $E(B-V)$ map (lower-left) and reconstruct the extragalactic LSS, or the CIB (upper), which we subtract to obtain CSFD, a cleaned dust map with improved accuracy (lower-right). The maps are in full-sky Galactic Mollweide projection with the Galactic center in the center and a histogram equalization color stretch. In the CIB map, gray areas mark regions without reconstruction (Galactic plane, Magellanic clouds, and M31), and the nontransparent areas indicate the most reliable regions for precision cosmology analyses. \textbf{Panel B:} Tomographic 1D (upper) and 2D (lower) statistics of the LSS imprints in SFD via cross-correlating with a set of reference galaxies and quasars in SDSS. The 1D estimator extracts the CIB-induced excess $E(B-V)$ within $10'$ of the references, and the 2D estimator measures the differential excess $E(B-V)$ at a given characteristic scale $\theta_C$ with a compensated filter, both as a function of redshift. \textbf{Panel C:} Same as panel B. but for CSFD, showing no detectable LSS imprint.}
    \label{fig:Summary}
\end{figure*}

If the goal is to get the pure Galactic component, e.g., a clean Milky Way dust map, we need to estimate the extragalactic component $I_{LSS}(\phi)$ and subtract it:
\begin{eqnarray}
    I_{G}(\phi) &=& I(\phi) - I_{LSS}(\phi), 
\label{eq:2}
\end{eqnarray}
as illustrated in Figure~\ref{fig:Summary}-A. For precise cleaning, the $I_{LSS}(\phi)$ here cannot be just a proxy, i.e., some map that correlates with the LSS;  it has to be a map-level reconstruction of the LSS component extracted from $I(\phi)$, the main task we are tackling in this paper. In the case of SFD, $I_{LSS}(\phi)$ is essentially the CIB at 100~$\rm \mu m$ but in an unintended unit of magnitude in $E(B-V)$. Although it is a contaminant in the Galactic reddening field, this CIB map, if successfully extracted, could be a valuable product in cosmology as a matter tracer. 

The challenges of this component separation problem are three-fold:
\begin{enumerate}
    \item The Galactic foreground is usually much brighter than the LSS signatures, which will add strong noise in estimating $I_{LSS}(\phi)$. In SFD, $I_{G}(\phi) > 100\,I_{LSS}(\phi)$ for most sight lines.    
    
    \item The extragalactic LSS component is a 2D projection of the 3D Universe: 
        \begin{equation}
        I_{LSS}(\phi) = \int \frac{dI_{LSS}}{dz}(\phi, z)\,dz, 
        \label{eq:3}
        \end{equation}
    where the radial information is lost. As the cosmic web is de-correlated at scales much greater than the correlation length, $\sim10$~Mpc or $dz \sim 10^{-2.5}$, the full LSS statistics should be sampled tomographically in sufficiently small redshift bins.
        
    \item Finally, unlike most inference problems in cosmology where the LSS is treated as a random field, a map-level reconstruction of $I_{LSS}(\phi)$ requires us to know all of the orientations or Fourier phases of the cosmic web (at least implicitly). This makes this problem intrinsically very high-dimensional. 
\end{enumerate}
Despite these difficulties, we lay out a novel component separation technique and apply it to clean SFD. Our method consists of two steps: first, to extract summary statistics of $I_{LSS}$ using tomographic cross-correlations, and second, to reconstruct, on the map level, $I_{LSS}(\phi)$ using a large set of empirical templates given the target statistics. Here we describe each step.

\subsection{Summary Statistics} \label{sec:formalism-stats}

The first step is to measure and to try to exhaust all detectable summary statistics of the $I_{LSS}$ field. These include, ideally, its power spectrum, bispectrum, and higher-order moments. In this work, we consider up to two-point statistics, as those beyond would be hard to measure in SFD (as the noise is raised to a higher power) and thus unlikely to cause systematics in cosmology at a detectable level in the near future. In reality, we do not have direct access to $I_{LSS}(\phi)$ but only the total map $I(\phi)$, which is dominated by $I_{G}(\phi)$. We, therefore, have to isolate the $I_{LSS}$ statistics via cross-correlations. 

The purpose of these cross-correlation-based summary statistics is two-fold. First, the amplitudes measured for a map $I(\phi)$ provide us with a test to see if the map is contaminated (or not) by the extragalactic LSS (e.g., Figure~\ref{fig:Summary}-B and C). Secondly, they are the ``target statistics'' we aim to reproduce in the reconstructed LSS field before subtracting it out to clean the original map $I(\phi)$.

\subsubsection{A Generic Tomographic Estimator} \label{sec:generic_estimator}
To isolate the LSS signatures on top of the foreground in the map $I(\phi)$ using cross-correlations, we need a reference sample $R$ of strictly extragalactic objects. This is usually from galaxies or quasars in well-tested LSS catalogs. As described earlier in challenge 2, if we aim to exhaust most LSS information, a subpercent level of redshift precision is needed to sample independent slices of the cosmic web along the line of sight. We thus consider using spectroscopic redshifts. Once the extragalactic reference $R$ is defined, we can access its 3D density contrast as a function of the 3D map $n_R(\phi, z)$ and its angle-averaged redshift distribution $\langle n_R(z)\rangle$:
\begin{eqnarray}
\delta_R(\phi, z) \equiv n_R(\phi, z)/ \langle n_R(z) \rangle -1. 
\label{eq:4}
\end{eqnarray}
This serves as a tracer of the underlying matter density field in the Universe and is fully de-correlated from the Galactic foreground.

We can write a generic redshift tomographic angular cross-correlation function in the configuration space: 
\begin{eqnarray}
w_{LSS, R}(\theta, z)\  &=&\  \langle \delta_R(\phi, z) \cdot \Delta I(\phi+\theta) \rangle \nonumber \\
\ &=&\ \langle \delta_R(\phi, z) \cdot \Delta I_{LSS}(\phi+\theta) \rangle,
\label{eq:5}
\end{eqnarray}
where $\Delta I(\phi) = I(\phi) - \langle I \rangle$, i.e., the mean removed or high-pass filtered field of the map $I(\phi)$, as the large-scale fluctuations are usually foreground dominated and need to be mitigated. Unlike the usual density contrast $\delta \equiv (n - \langle n \rangle)/\langle n \rangle$, here $\Delta I$ is not normalized as it is both unnecessary and not doable under the presence of foregrounds. One can appreciate the key feature of this estimator: although the input map $\Delta I(\phi)$ contains all line-of-sight components (Equation~\ref{eq:1}), non-LSS terms drop out as the reference $R$ is purely extragalactic. If a wide redshift range and angular scale are covered, with sufficient signal-to-noise ratios (SNR), all information on the two-point cross LSS statistics would be extracted.

\subsubsection{Comparison to Clustering Redshift Estimation} \label{sec:clustering_z}
This redshift-dependent cross-correlation analysis is similar to the so-called clustering redshift estimation where one can tomographically extract the redshift distributions $P(z)$, or in a more general case, the bias-weighted equivalent $b(z)P(z)$ for an arbitrary source population \citep[][and see \href{http://tomographer.org/}{\textit{Tomographer}},\footnote{\url{http://tomographer.org/}; Chiang et al., in prep.} for a generic tool]{2008ApJ...684...88N, 2013MNRAS.433.2857M, 2013arXiv1303.4722M}. In fact, the $w_{LSS, R}(\theta, z)$ estimator in Equation~\ref{eq:5} takes the  same form as one of the steps in the general clustering redshift formalism as laid out in \cite{2019ApJ...877..150C} for diffuse intensity maps. 

There are, however, a few differences between the tomographic cross-correlations in this work and clustering redshift analyses. First, the target $P(z)$ in clustering redshift is a scale-independent quantity; the estimator used there is thus marginalized over the angular dimension. In this paper, however, since the purpose is to characterize and reconstruct the full, highly nonlinear LSS field, we are interested in the angular dependence of $w_{LSS, R}(\theta, z)$ all the way into the smallest scales. Secondly, in this work, we do not need to post-process $w_{LSS, R}$ to get a proper redshift probability function; if one were to do so, corrections for matter clustering and the bias factor for the reference objects would be needed \citep{2013arXiv1303.4722M}.

\subsection{Reconstruction} \label{sec:formalism-recon}

Having measured an exhaustive set of tomographic statistics via cross-correlations (Equation~\ref{eq:5}), we consider reconstructing the LSS component $I_{LSS}(\phi)$ on the map level. As mentioned previously in challenge 3, the phases of the LSS on our observable sky are not random and need to be specified. As no cosmology theory can provide such information, we consider a data-driven approach using a large set of empirical templates, which we call $T_{LSS, i}(\phi)$, where $i$ specifies the template number. 

The LSS templates $T_{LSS, i}(\phi)$ are usually each a sky map of density contrast built from some generic extragalactic tracers, i.e., galaxies or active galactic nuclei (AGNs). There are two essential requirements. First, to form a complete ``basis set'' for reconstructing the LSS, $T_{LSS, i}(\phi)$ have to be from sufficiently diverse populations of objects in terms of redshift coverages and clustering properties. Secondly, to avoid introducing new foregrounds, the templates $T_{LSS, i}(\phi)$ should be clean and purely extragalactic; this makes it impractical to consider diffuse intensity maps (e.g., CO, [CII], HI 21~cm, or $Planck$ CIB maps) as LSS templates. 
When building the templates $T_{LSS, i}(\phi)$, we do not need to start with sources with known redshifts. This means that data for the template sources are not as demanding as the reference, and the reconstruction should be feasible over most of the extragalactic sky with already existing broadband surveys as the templates.

Once we build a complete basis template set that encodes the possible space of orientations or phases of the LSS, we need to specify the rule to combine them. The simplest model is to require the reconstructed LSS field to be a linear combination of the basis templates:
\begin{equation}
I_{LSS}^{\,rec}(\phi) = \sum_{i=0}^{N}\,C_i\,T_{LSS, i}(\phi),
\label{eq:6}
\end{equation}
where $C_i$ is the linear coefficient, or weight, for each of the $N+1$ templates. For most templates, we do not know the  $C_i$ to begin with, while for a small subset, strong priors might be available. For example, in the case of the CIB in SFD, one can treat a sky map constructed from known but previously unmasked FIR bright galaxies as a point-source template, where the weight $C_i$ can be set readily by their FIR fluxes.

As a core concept of our reconstruction, to determine the full set of $C_i$ for all templates, including the majority with no prior information, we minimize
\begin{equation}
\| w_{LSS, R}^{\,rec}(\theta, z) - w_{LSS, R}(\theta, z)\| , 
\label{eq:7}
\end{equation}
the difference between the tomographic statistics for the reconstructed field and the targeted ones measured for $I_{LSS}$ in the original map $I$. Here we use the L2 norm, i.e., the root mean square difference between the two $w$ vectors. As we use a linear model for reconstruction, the two-point cross-statistics also combine linearly:
\begin{eqnarray}
w_{LSS, R}^{\,rec}(\theta, z)\ &=&\ \langle \delta_R(\phi, z) \cdot \Delta I_{LSS}^{\,rec}(\phi+\theta)\rangle \nonumber \\
&=&\ \sum_{i=0}^{N}\,C_i\,\langle \delta_R(\phi, z) \cdot \Delta T_{LSS, i}(\phi+\theta)\rangle \nonumber \\
&=&\ \sum_{i=0}^{N}\,C_i\,w_{TR, i}(\theta, z),
\label{eq:8}
\end{eqnarray}
where the estimator in Equation~\ref{eq:5} is now applied to every template $T_{LSS, i}(\phi)$ to get $w_{TR, i}(\theta, z)$ in exactly the same way (and the same redshift and angular bins) as $w_{LSS, R}(\theta, z)$ for the original map $I(\phi)$. It is at this stage we gain redshift information for each of the LSS templates through tomographic cross-correlations.

To summarize, the reconstructed LSS field $I_{LSS}^{\,rec}(\phi)$ in any sky map $I(\phi)$ is a linear combination of templates $T_{LSS, i}(\phi)$ (Equation~\ref{eq:6}). These templates are made from known extragalactic objects with sufficient diversity to cover the possible phase-space of the LSS. The linear coefficients are then determined such that the cross-correlation-based LSS statistics (Equation~\ref{eq:5}) of the reconstructed field $I_{LSS}^{\,rec}(\phi)$ is indistinguishable from the target statistics measured in $I(\phi)$. 

After obtaining the reconstructed LSS field, we can simply subtract it out to get an estimate of the cleaned, LSS-free Galactic map: 
\begin{equation}
\hat{I}_{G}(\phi) = I(\phi) - I_{LSS}^{\,rec}(\phi).
\label{eq:9}
\end{equation}
Compared to Equation~\ref{eq:2}, Equation~\ref{eq:9} acknowledges that there could be errors in the reconstruction and, thus, the LSS cleaning. In practice, it is important to make attempts to maximize the SNR for the reconstruction and to validate the results independently.

\section{SFD Component Separation} \label{sec:comp_sep_SFD}

Having laid out the component separation method, we apply it to clean the SFD dust reddening map, formally, $I(\phi) = E(B-V)_{SFD}(\phi)$. Equation~\ref{eq:1} then becomes
\begin{eqnarray}
&E&(B-V)_{SFD}(\phi)\, = \, \nonumber \\
&E&(B-V)_{CSFD}(\phi)\, + \, E(B-V)_{CIB}(\phi), 
\label{eq:10}
\end{eqnarray}
where the pure Galactic component would be what we call CSFD, the corrected SFD, and the extragalactic LSS component is essentially the 100~$\mu$m CIB in the reddening unit. Our goal is then to estimate the CIB term via reconstruction: $E(B-V)_{CIB}(\phi) = I_{LSS}^{rec}(\phi) + error(\phi)$.

As dust map cleaning is among the most challenging cases where the foreground is much stronger than the LSS signatures, in every step of our analysis, we pursue aggressive optimization to achieve maximal SNRs. In addition, this work is built on the synergy between a rich set of multiwavelength sky survey catalogs. We thus defer a substantial amount of technical information to the Appendix. Here we briefly describe our data and analysis before presenting the resulting products. Throughout this section, if not specified otherwise, all maps are handled using the HEALPix \citep{2005ApJ...622..759G} scheme with $N_{side}=2048$ (pixel size 1.72') to sufficiently Nyquist sample the $6.1'$ beam set by SFD. 

For the reference sample $R$ used in extragalactic cross-correlations, we compile 2.7 million galaxies and quasars with spectroscopic redshifts up to $z\sim 3$ from seven well-crafted LSS catalogs in SDSS \citep[][Appendix~\ref{sec:ref}]{2000AJ....120.1579Y}. For the LSS templates $T_{LSS, i}(\phi)$ used in the field-level CIB reconstruction, we build 180 sky maps of density contrasts from 604 million galaxies in Wide-field Infrared Survey Explorer \citep[WISE;][]{2010AJ....140.1868W} from the CatWISE2020 \citep{2021ApJS..253....8M} and WISE$\times$SuperCOSMOS \citep{2016ApJS..225....5B} catalogs. We split these galaxies in either the color-magnitude or photometric-redshift space to maximize the diversity and, thus, the completeness of the LSS templates (Appendix~\ref{apd:base_templates}). These features (color, magnitude, and photometric redshift as a dimensionality-reduced color index) are merely labels used to separate different populations of galaxies, while once the templates are built, our LSS reconstruction uses only their tomographic clustering information (Equations~\ref{eq:6}-\ref{eq:8}). As a crucial step, templates built from each of these WISE subsamples are further augmented to have a range of beam sizes from sub- to super-SFD resolutions (Appendix~\ref{sec:augmentation}). On large, linear scales, templates beam-augmented from a given WISE subsample are identical, while on small scales, the combination allows us to mimic a wide range of possible nonlinear clustering of the CIB at the redshift range sampled by each WISE population.

After the IRAS mission and the release of SFD, some more dusty galaxies over the full sky have been identified directly in the FIR by the AKARI satellite \citep{2007PASJ...59S.369M}. As a supplement to the main, 180 WISE-based LSS templates, we add one additional AKARI point-source template as the zeroth $T_{LSS}(\phi)$, whose contribution to the CIB in SFD can be fixed by calibrating the stacked FIR flux to $E(B-V)$ relation, effectively masking these sources out in cleaning SFD (Appendix~\ref{sec:akari}).

Although the template data are full-sky, we split the sky area into three tiers, as shown in the reconstructed LSS/CIB map in Figure~\ref{fig:Summary}-A (upper). The gray marks the Galactic plane and bulge, Magellanic Clouds, and M31, where we make no attempt to reconstruct the LSS as the quality of the templates is largely degraded. Within our footprint, the transparent areas indicate where the reconstruction is less reliable (extended footprint boundaries and regions strongly affected by dust cirrus, bright stars, and clusters; see Appendix~\ref{apd:base_templates}). The nontransparent field indicates the ``cosmology area''  for precision cosmology analyses, covering 90\% of the high-latitude sky at $|b|>30^{\circ}$ (or 58\% full sky). Independent from these three tiers, there is 2\% of the sky with no IRAS coverage, where SFD is based on the lower-resolution (1$^{\circ}$) DIRBE data. We exclude these DIRBE-only regions from our cross-correlations, while for the part within the footprint, the SFD cleaning (Equation~\ref{eq:10}) is done using the reconstructed CIB smoothed with a matched, 1$^{\circ}$ beam.

Having the $E(B-V)(\phi)$ field in SFD and the 180 WISE plus one AKARI LSS templates $T_{LSS, i}(\phi)$, we then use the SDSS references $R$ to extract extragalactic statistics via cross-correlations in all these maps before the CIB reconstruction (minimizing Equation~\ref{eq:7}) can proceed. Instead of using the redshift-angular cross-correlation function (Equation~\ref{eq:5}), the estimators we end up applying in this work are two derived variants: a 1D estimator optimized for redshift tomography and a 2D estimator optimized for extracting angular information. The two combined carry the same information as Equation~\ref{eq:5} but are less subjected to foreground-induced bias and have cleaner covariances. Below we define these two estimators and show the measurements for SFD; the results for the $180+1$ LSS templates are shown in Appendix~\ref{sec:extract_stats_temp}.

\subsection{1D Information} \label{sec:extract_stats_SFD_1D}
Starting from $w_{LSS, R}$ in Equation~\ref{eq:5}, we derive the 1D estimator optimized for redshift tomography. In this case, the angular dependence can be integrated out:
\begin{eqnarray}
w_{LSS, R}^{1D}(z)\  &=&\  \int_0^{\theta_{max}} W(\theta)\, w_{LSS, R}(\theta, z)\,d\theta,
\label{eq:11}
\end{eqnarray}
where $W(\theta)$ is an arbitrary weight function. Here we choose to set $W\propto \theta$ such that $w_{LSS, R}^{1D}(z)$ is simply an aperture photometry measured around the SDSS references. We set the integration bound $\theta_{max} = 10'$ to capture the regime with strong clustering signals. Since the contribution over some angular scales is combined, we can afford a dense sampling scheme along the $z$-dimension with 64 bins in $log(1+z)$ from $z=0$ to $3$ while still staying in the signal-dominated regime. Applying to SFD, this estimator $w_{LSS, R}^{1D}(z)$ becomes $\Delta E(B-V)_{10'}(z)$, the excess reddening within $10'$ of the SDSS references. The measurements are shown in the upper panel in Figure~\ref{fig:Summary}-B, where we find that the 100~$\mu$m CIB in SFD is peaking at $z<0.1$ with an amplitude of $\sim0.2$ mmag; meanwhile, the high-redshift tail extends up to $z\sim2.5$, similar to the result found already in CM19. Here we are able to achieve high SNRs even under the presence of the strong Galactic foreground thanks to an HI-based foreground mitigation, filtering, and optimal weighting scheme explained in Appendix~\ref{apd:foreground}. 

We note that the actual $P(z)$ of the CIB photons, or $dE(B-V)/dz(z)$ for SFD would be peaking at a higher redshift ($\sim0.5$--$1$) based on our run of a proper clustering redshift analysis using \href{http://tomographer.org/}{\textit{Tomographer}}. The main factor that drives the peak redshift of $\Delta E(B-V)_{10'}$ lower is the evolution of the underlying matter clustering $w_m(z)$, which is rising with decreasing redshift. We do not correct for $w_m$ (nor the galaxy bias for the SDSS references) in this work as the exact same factor will also enter the statistics of the LSS templates and becomes only a normalization factor in Equation~\ref{eq:7}, therefore not affecting the LSS reconstruction. We apply this 1D estimator to also measure the statistics for all of the $180+1$ LSS templates $T_{LSS, i}(\phi)$, with results shown in Appendix~\ref{sec:extract_stats_temp}.

The error bars in Figure~\ref{fig:Summary}-B are obtained by resampling the SFD map using a procedure extended from that introduced in \cite{2008ApJ...681..726L}.
We create 200 bootstrap realizations of the sky by resampling spatial blocks with replacements. The blocks are defined using HEALPix $N_{side}=16$ pixels with a size (3.66~deg)$^2$ each, which is chosen such that the data are spatially de-correlated and independent among blocks. We keep the same set of block-bootstrap realizations throughout the paper when we measure the extragalactic signatures of different maps (SFD and the LSS templates). This way, the complex covariances between data vectors can be evaluated.

\subsection{2D Information} \label{sec:extract_stats_SFD_2D}

We now need a 2D estimator for extracting the angular dependence. In principle, the $w_{LSS, R}(\theta, z)$ in Equation~\ref{eq:5} can be used directly. However, to push for more robust measurements under the presence of the strong foreground in SFD, we show, in Appendix~\ref{apd:comp}, that it is more stable to use a compensated top-hat filter. This takes a differential form by simply subtracting a local zero-point measured immediately outside the aperture at scale $\theta_C$:
\begin{eqnarray}
w_{LSS, R}^{2D}(\theta_C, z)\  &=&\ \int_0^{\theta_C} W(\theta)\, w_{LSS, R}(\theta, z)\,d\theta  \nonumber \\
&-&\ \int_{\theta_C}^{\sqrt{2}\theta_C} W(\theta)\, w_{LSS, R}(\theta, z).
\label{eq:12}
\end{eqnarray}
Here we apply the same area-preserving weight $W\propto \theta$ as in the 1D case, and the factor of $\sqrt{2}$ in the bound sets the inner aperture and outer annulus equal area. Applying it to SFD, we denote this 2D quantity as $\Delta E(B-V)_{C}(\theta_C, z)$, which captures the slope, at the scale of $\theta_C$, of the biased $E(B-V)$ due to the LSS contamination in SFD. We show the measurements in the lower panel in Figure~\ref{fig:Summary}-B, where we use 16 $\theta_C$ bins logarithmically spaced over $\sqrt{2}'$ and $256'$. The redshift resolution is one-fourth of that in the 1D case to allow for higher SNRs in the angular dimension. We see that in SFD, the clustered CIB is strong at scales around the beam, while at low redshifts, the large-scale signal is also significant up to $\theta_C = 1$--2$^{\circ}$. At $z>1$, the large-scale contribution becomes too noisy to detect. In the same manner as the 1D case in Section~\ref{sec:extract_stats_SFD_1D}, the 2D measurements are done after applying the HI-based map-level cleaning to SFD (Appendix~\ref{apd:filtering_HI_clean}), and the errors and covariances are obtained using block-bootstrapping. We apply this 2D estimator to also measure the statistics for all of the $180+1$ LSS templates $T_{LSS, i}(\phi)$, the results of which are shown in Appendix~\ref{sec:extract_stats_temp}.

Despite the strong Galactic foreground, the cross-correlations between SFD reddening and SDSS references extract clean, foreground-free information on the LSS in SFD. The combined vector of the 1D plus 2D tomography has a high dimensionality of 320, providing a set of very constraining target statistics for a map-level reconstruction.

\subsection{Computation} \label{sec:computation}

Altogether, the number of cross-correlations we calculate in this work is huge and is dominated by extracting the statistics for the LSS templates: $N_{template} \times N_{data\ vector} \times N_{bootstrapping} = 181\times 320\times 200 = 11.6$ million. Such a demanding computation is made possible by a novel technique developed in \href{http://tomographer.org/}{\textit{Tomographer}} (Chiang et al., in prep). Briefly, utilizing the fact that one side of the cross-correlations is always fixed to the same set of reference galaxies, we pre-calculate the pairs between every reference object and every HEALPix pixel in the sky. We then store the binned pair counts as effective weights in giant look-up tables. After recasting the test data (e.g., SFD or LSS templates) into HEALPix arrays with the pre-defined format, each cross-correlation is then a simple vector multiplication with no actual spatial pair finding occurring at run-time. With this algorithm, we can achieve the best possible scaling for the computing time with $\mathcal{O}(1)$, i.e., constant. This is to be compared with the usual $\mathcal{O}(N^2)$ scaling in typical calculations of two-point statistics where the computing time scales vary rapidly with $N$, the number of spatial resolution elements or objects involved. With this algorithm, we are able to finish the computation in this work with reasonable resources (about 200~GB storage and 0.1~million CPU hours, with a large fraction used for testing).

\begin{figure*}[t!]
    \begin{center}
         \includegraphics[width=1\textwidth]{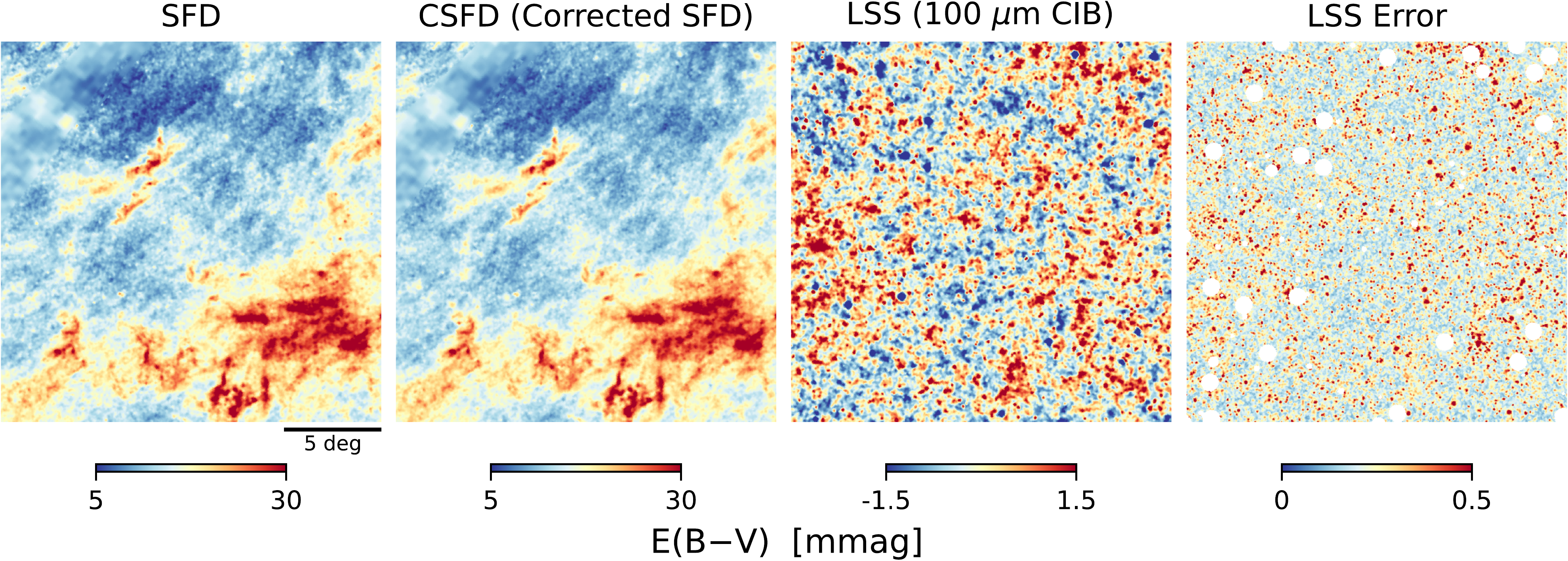}
    \end{center}
    \caption{Comparison between SFD, the corrected SFD (CSFD), the LSS correction term, and its error from left to right panels, respectively, over a 20$^{\circ}$ field centered on the Galactic coordinates $(l, b) = (-50^{\circ}, -80^{\circ})$ in Gnomonic projection. All maps are in the SFD $E(B-V)$ unit, with a decreasing dynamic range from the left to right panels. The CSFD map is made to be LSS-free, i.e., $\mathrm{CSFD} = \mathrm{SFD} - \mathrm{CIB}$. The white masks in the last panel, mostly around bright, red stars in this field, indicate the ``noncosmology'' areas where the LSS reconstruction, and thus SFD correction, is less reliable (Appendix~\ref{apd:base_templates}).}
    \label{fig:Gnomonic_zoom}
\end{figure*}

\subsection{Reconstruction in Practice} \label{sec:reconstruction}

We are now ready to reconstruct the full-sky LSS or the 100~$\rm \mu m$ CIB map embedded in SFD using the statistics obtained by cross-correlations and the Fourier phases implicitly provided by the templates $T_{LSS, i}(\phi)$. 

We seek a linear combination of the templates such that the reconstructed map has indistinguishable tomographic statistics from the CIB in SFD. This is equivalent to fitting the linear coefficients $C_i$, or weights, in Equation~\ref{eq:6}, one for each template under the constraints in Equations~\ref{eq:7} and \ref{eq:8}. During the fitting, we fix only $C_0$ for the $0$th template from AKARI galaxies as they are directly detected in the FIR (Appendix~\ref{sec:akari}). Since the dimensionality of the $C_i$ posterior space is high (180), it would not be feasible, computationally, to proceed with the standard Markov Chain Monte Carlo (MCMC) approach. Furthermore, it is unclear how the covariances of, e.g., the templates, should be incorporated in MCMC. For these reasons, we design a fitting procedure that is practical yet outperforms MCMC in terms of error handling, taking advantage of our spatial resampling with block-bootstrapping. 

In the $m$-th bootstrap realization, we proceed with an end-to-end inference all the way to create a reconstructed LSS map ${I_{LSS}}^m(\phi)$. This is achieved by fitting the set of weights ${C_i}^m$ under the same constraint in Equations~\ref{eq:7} and \ref{eq:8} but now with tomographic statistics for SFD and the templates measured in this specific, $m$-th bootstrap realization. After going over all the $N_{bootstrapping} = 200$ realizations, we then have 200 reconstructed LSS maps ${I_{LSS}}^m(\phi)$. This way, we can examine the posterior space on the map level directly. For each sky pixel (HEALPix $N_{side}=2048$), we evaluate the statistics over the 200 samplings. Collecting these values across the sky, we finally arrive at our best estimate for the reconstructed LSS map $I_{LSS}^{rec}(\phi)$ at $6.1'$ resolution using the posterior medium, as well as the associated error map using posterior standard deviation. As expected, most of the LSS fluctuations are reconstructed from the overcomplete set of WISE templates, with only the tip of the CIB resolved in AKARI.

To clean SFD from extragalactic contamination, we simply subtract the reconstructed LSS map in the extinction unit (rearranging Equation~\ref{eq:10}) to obtain CSFD---the corrected SFD Galactic dust reddening map. In the small fraction of the sky where SFD is of lower, 1$^{\circ}$ resolution, the LSS field is further smoothed to 1$^{\circ}$ for the correction. In the Galactic plane, Magellanic Clouds, and M31 areas (gray in Figure~\ref{fig:Summary}-A; upper) where no attempt on the LSS reconstruction is made, we keep the original SFD $E(B-V)$ values in CSFD. To this end, we have completed the SFD Galactic-extragalactic component separation using a data-intensive, maximally model-independent approach.

To facilitate the discussion on our map products in the following sections, we can also convert the unit of the reconstructed LSS/CIB map $I_{LSS}^{rec}$ and the associated errors from magnitude in $E(B-V)$ back to the usual intensity unit, e.g., $\rm MJy\ sr^{-1}$. This can be proceeded by reversing the conversion adapted in \cite{1998ApJ...500..525S}, $E(B-V) = p\,I_{\rm IRAS\ 100\mu m}$ with $p = 0.0184$ and $I_{\rm IRAS\ 100\mu m}$ is the temperature-corrected 100~$\mu$m flux. In what follows, we will use these two units interchangeably.

\subsection{CIB and CSFD Map Products} \label{sec:maps}

We now examine our new map products, the reconstructed CIB and the pure Galactic CSFD dust reddening maps. To appreciate the main features, we first visualize and compare the original SFD, the new CSFD, the reconstructed LSS, and the associated LSS error map over a 20-deg scale in Figure~\ref{fig:Gnomonic_zoom} from left to right, respectively. This field (centered on Galactic $l = -50^{\circ}$; $b=-80^{\circ}$) is chosen to be representative of a typical high-latitude, low dust column density patch of sky with a few clouds of Galactic cirrus. Visually, it is hard to distinguish CSFD (second panel) from SFD (first panel), as our LSS cleaning, by construction, preserves the large-scale textures of the dust cirrus in the Milky Way ISM. 

The reconstructed LSS (third panel) exhibits a homogeneous and isotropic field with fluctuations on mostly small scales, consistent with what one expects for a cosmological density field. Although the locations of the over- and underdense spots appear random, they actually track, faithfully, the cosmic web in the Universe we live in, with the phase information coming from our empirical LSS templates from real galaxy density fields. In the last panel, the error map also appears homogeneous and isotropic, at least on the 20$^{\circ}$ scale. We note that the dynamic range in $E(B-V)$ in Figure~\ref{fig:Gnomonic_zoom} decreases from the left two to the right, which echoes the following picture: the LSS contamination is a relatively small bias or a correction term from SFD to CSFD, and the error of this bias (fourth panel) is even smaller, suggesting that the reconstruction is signal-dominated. Although the LSS term in SFD is small, the fact that we can detect it at a high SNR means that it will become an important systematic in precision cosmology and astrophysics in the foreseeable future, if not already now.

We now zoom out to examine the global behaviors of our map products. We show the full-sky SFD, CSFD, and the reconstructed CIB map already in Figure~\ref{fig:Summary}-A. As in the smaller-scale picture in Figure~\ref{fig:Gnomonic_zoom}, our new CSFD looks very similar to the original SFD globally. This is because CSFD, by construction, preserves all the Milky Way ISM features. At high latitudes, both dust maps have a typical $E(B-V)$ of about 20 mmag. The CSFD, however, would contain minimal extragalactic contamination, which we will validate further. The reconstructed LSS map shown on top of Figure~\ref{fig:Summary}-A, again, looks like a homogeneous random field, while the amplitudes, locations, and orientations of the LSS, including clusters, filaments, and voids, are, by construction, faithful reproduction of that in our Universe. There is no appreciable gradient along Galactic latitude, consistent with what we expect for a foreground-free map. 

\begin{figure}[t!]
    \begin{center}
         \includegraphics[width=0.48\textwidth]{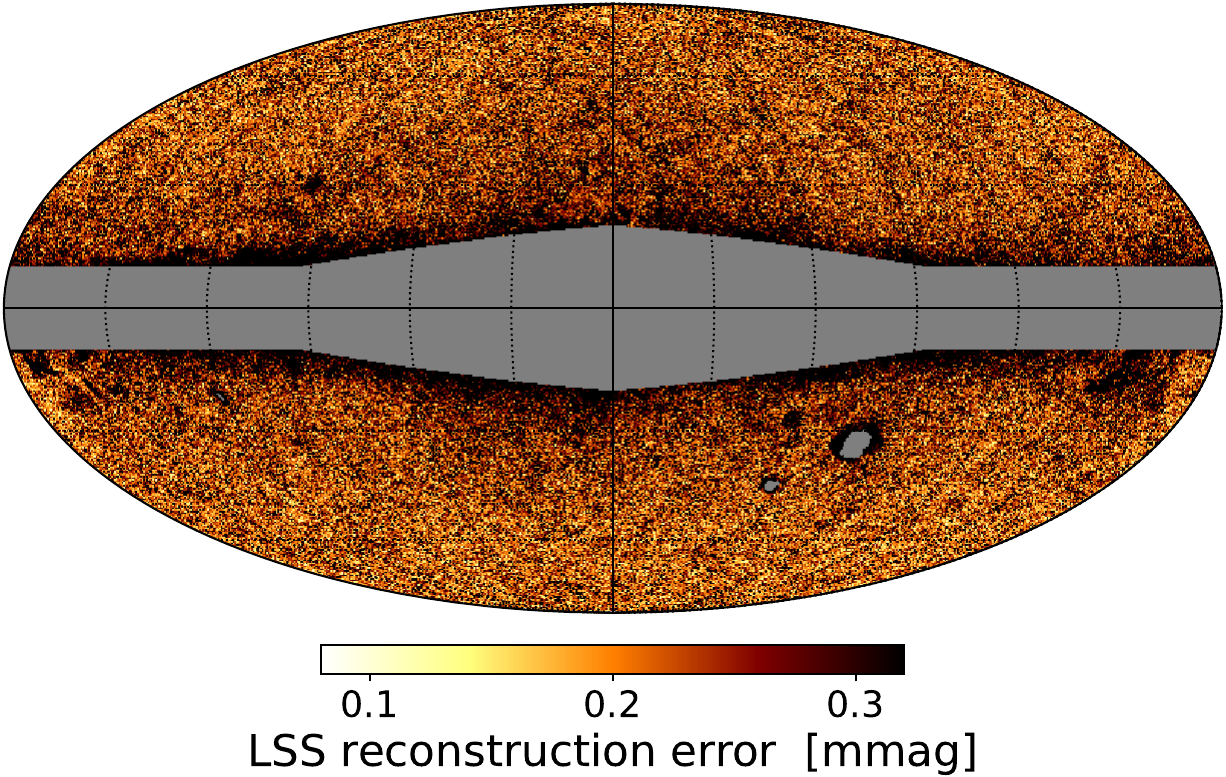}
         
    \end{center}
    \caption{Error map of the LSS/CIB reconstruction in the SFD reddening unit in full-sky Galactic Mollweide projection. The typical error is 0.2 mmag and slightly increases toward low latitudes due to increasing uncertainties in mostly the WISE-based LSS templates.}
    \label{fig:rec_error_map}
\end{figure}

In Figure~\ref{fig:rec_error_map}, we further visualize the full-sky error map for the LSS reconstruction (zoom out from the last panel of Figure~\ref{fig:Gnomonic_zoom}). The full reconstruction area is shown , while the errors are more reliable in the unmasked, cosmology area (nontransparent part of the reconstructed LSS/CIB map in Figure~\ref{fig:Summary}-A). In this error map, we see an increasing noise toward low Galactic latitudes, a feature reassuring that our fitting procedure in Section~\ref{sec:reconstruction} can propagate uncertainties and covariances inherent in the data products we use (SFD, the SDSS reference, and the LSS templates). We identify that the main reason for the latitude-dependent noise is from our WISE-based LSS templates (Appendix~\ref{sec:templates}), whose quality degrades toward low-latitude areas as the photometric errors due to both extinction and crowding increase.

\begin{figure}[t!]
    \begin{center}
         \includegraphics[width=0.47\textwidth]{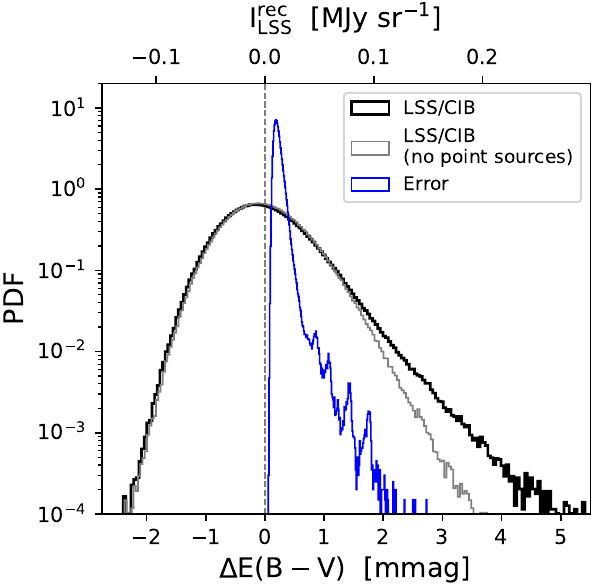}
    \end{center}
    \caption{Probability distributions of the per-beam, or per-resolution, element map values for the best reconstructed LSS map (black, and that without the AKARI point-source contribution in gray) and its corresponding error map (blue) in our cosmology area. Independent resolution elements are approximated using HEALPix $N_{side}=512$ pixels of $6.9'$ each, similar to the beam size of SFD. The typical LSS correction is at the millimagnitude level, and the mean SNR is $\sim2.5$ for every resolution element.}
    \label{fig:per_beam_pdf}
\end{figure}

We now quantify the SNR of our reconstructed LSS/CIB map, which is also the correction term to derive CSFD. To incorporate correlated behaviors within the beam, we rebin, on the individual bootstrap realization level, the reconstructed LSS and error maps to HEALPix $N_{side}=512$ resolution with $6.9'$ pixels to approximate the resolution elements of SFD. With that, nearby pixels should not be strongly correlated due to over-sampling. Figure~\ref{fig:per_beam_pdf} shows the probability distributions for the per-beam map values for the reconstructed LSS (black), LSS without the AKARI point-source contribution (gray), and the error of the LSS (blue) in the cosmology area.\footnote{Plotting the same for the over-sampled $N_{side}=2048$ maps would result in very similar but slightly wider distributions.} The LSS map has both positive and negative values, with a mean very close to zero by construction (i.e., there is no monopole) as we do not intend to alter the large-scale zero-point of SFD. In this double-sided case, the typical amplitude of the LSS correction can be approximated by its standard deviation, which is 0.7 mmag per beam. At high latitudes, this corresponds to a 3.5$\%$ correction to the original SFD, with the tail going as high as a 40$\%$ correction for a small number of sight lines. Over the footprint of our reconstruction, most high CIB sight lines coincide with local, massive galaxy clusters, including Abell 2256, 539, 2319, 3266, 3158, 2255, 3744, 3558, 2065, and Coma, while some are already marked as noncosmology areas due to exactly these overdensities and crowding in WISE data.

Due to the zero-mean nature of the LSS fluctuations, it is not straightforward to estimate the per-beam SNR. Here we give two examples of such estimates in the unmasked cosmology area where the reconstruction is most reliable. First, we consider dividing the typical signal, i.e., the standard deviation of the LSS term, by the mean error: $SNR_{per\ beam} \approx \sigma(I_{LSS,i}^{\,rec}) / \langle \sigma_{LSS,i}^{\,rec} \rangle = 2.9$. Alternatively, we can simply take the mean of the per-pixel absolute correction over its error: $SNR_{per\ beam} \approx \langle |I_{LSS,i}^{\,rec}| / \sigma_{LSS,i}^{\,rec} \rangle = 2.5$. Given that we have over 2 million quasi-independent resolution elements within the unmasked area over the sky, the combined SNR should be easily of the order of thousands. Our reconstructed LSS map, and therefore the SFD correction, are thus highly signal-dominated.

\section{Validation} \label{sec:validation}
Here we validate our component separation of SFD into CIB plus CSFD. If the analysis is effective, our reconstructed CIB map should be foreground-free, and the CSFD dust map should be LSS-free.

\subsection{How Good Is the CIB Map?} \label{sec:CIB-iras_100micron}

Our reconstructed 100~$\mu$m LSS map $I_{LSS}^{rec}(\phi)$ is a potentially valuable cosmology product by itself as a matter tracer in the Universe. Here we ask two questions: (1) How well does it trace the extragalactic LSS? and (2) Is it truly foreground-free?

\begin{figure}[t!]
    \begin{center}
         \includegraphics[width=0.47\textwidth]{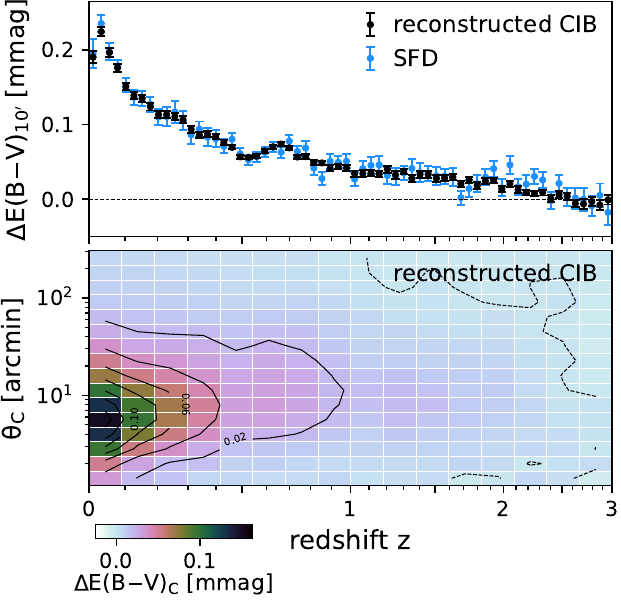}
    \end{center}
    \vspace{-1.1mm}
    \caption{Tomographic cross-correlation statistics, same as Figure~\ref{fig:Summary}-B and C, but for the reconstructed LSS/CIB map. The 1D statistics for SFD (from Figure~\ref{fig:Summary}-B) are overlaid using blue data points in the upper panel. By construction, the reconstructed CIB map reproduces all statistics of the CIB embedded in SFD. Furthermore, the SNRs in the former are much higher, highlighting the value of this foreground-free LSS product as a probe of the matter density field in the Universe.}
    \label{fig:rec_LSS_x_SDSS}
\end{figure}

For the first question, in addition to the high combined SNR demonstrated in Section~\ref{sec:maps}, one check we can do is to measure the tomographic LSS statistics by feeding the reconstructed CIB map into our 1D plus 2D cross-correlation pipeline (Sections~\ref{sec:extract_stats_SFD_1D} and \ref{sec:extract_stats_SFD_2D} with spectroscopic SDSS references). Figure~\ref{fig:rec_LSS_x_SDSS} shows exactly these statistics. The 1D ones measured for the original SFD map (from Figure~\ref{fig:Summary}-B) are also overlaid in the top panel with blue data points. By construction, our reconstructed CIB map reproduces all tomographic LSS statistics in SFD. Encouragingly, we see a much lower level of noise in these cross-correlations for the reconstructed CIB map (black data points) than SFD (blue) judging by both the error bars and the fluctuations of these data points over redshift. This is consistent with our expectation that once the foreground is ``stripped'' on the map level, our CIB map could be a better, cleaner, and yet unbiased product of cosmological matter tracer in the Universe.

\begin{figure}[t!]
    \begin{center}
         \includegraphics[width=0.47\textwidth]{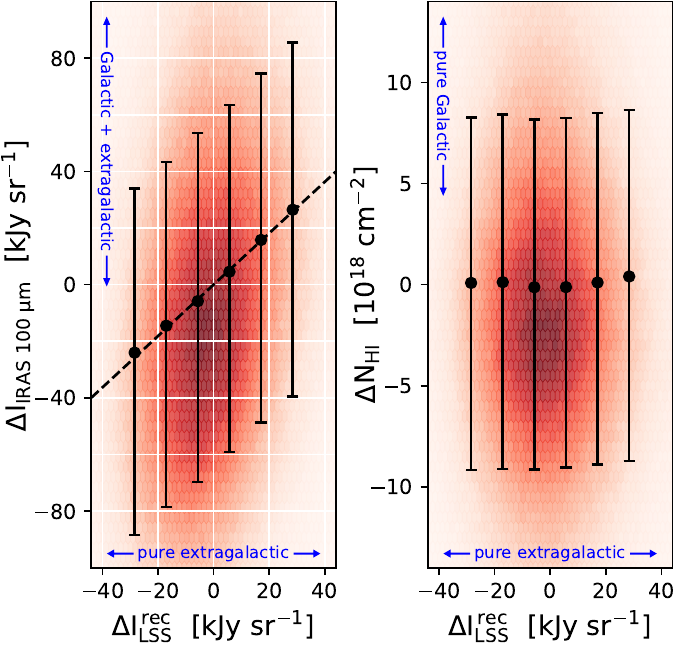}
    \end{center}
    \caption{Pixel value correlations between our reconstructed LSS/CIB map $\Delta I_{LSS}^{rec}$ on the $x$-axis versus the IRAS 100~$\mu$m channel map ($\Delta I_{\rm IRAS\ 100\ \mu m}$; left panel), and the Galactic HI map ($\Delta N_{\rm HI}$; right panel) on the $y$-axis over the 5\% of the sky with the lowest foreground. All maps are beam-matched and high-pass filtered over 1$^{\circ}$. Data points show the binned averages and 68th percentile ranges. The $\Delta I_{LSS}^{rec}$ correlates with $\Delta I_{\rm IRAS\ 100\ \mu m}$ (left) with a best-fit linear relation (dashed line) close to an unbiased, one-to-one relation. The scatter in $\Delta I_{LSS}^{rec}$ ($x$-axis) is much smaller, as, unlike the IRAS channel map, our LSS map is foreground-free, which can be verified by the lack of correlation seen with the purely Galactic HI in the right panel. }
    \label{fig:vs_IRAS100_NHI}
\end{figure}

We can also compare our reconstructed LSS/CIB map in the intensity unit with the IRAS 100~$\mu$m channel map \citep{2005ApJS..157..302M}, as both should trace the 100~$\mu$m CIB (while the latter contains foreground, just like SFD). In addition, to address the second question that if our reconstructed CIB map is truly foreground-free, we can compare it with the spectroscopic HI 21~cm map from the HI4PI survey \citep{2016A&A...594A.116H}, which has been tested to be free of extragalactic contamination in CM19. In Figure~\ref{fig:vs_IRAS100_NHI}, we show the 2D histograms for the pixel values (HEALPix $N_{side}=2048$) in our reconstructed LSS map versus the IRAS 100~$\mu$m channel map on the left panel and versus the Galactic HI column density (within $|v\rm_{LSR}| < 90\ km\ s^{-1}$) on the right panel. We use a large aspect ratio intentionally, as the scatter in our reconstructed LSS map (x-axis) is much smaller because of the low noise. In this plot, pixel values from all three maps are beam-matched to $16.1'$ set by HI. The sky area is restricted to the $5\%$ cleanest high-latitude sky with the lowest Galactic dust emission; additional high-pass filtering with 1$^{\circ}$ FWHM is applied to suppress the foreground along the y-axis further. To better examine the correlation (or not), we also show the binned averages and scatters using the data points. In the left panel, we see that our reconstructed LSS map indeed positively correlates with the IRAS 100~$\mu$m channel map, with a significant Pearson correlation coefficient of $0.20$. Again, we do not expect a perfect correlation as the IRAS 100~$\mu$m map is subject to foreground contamination. Nonetheless, it is worth pointing out that the best-fit linear relation (dashed line) is close to a one-to-one relation (i.e., 45$^{\circ}$ in this panel with the same unit in both axes), with a slope of 0.91. We find that this small deviation of the slope from unity can be attributed to the mean temperature correction done in SFD. This provides additional validation that our reconstructed LSS/CIB map is a robust tracer of the 100~$\mu$m CIB with a slightly modulated normalization. 

In the right panel in Figure~\ref{fig:vs_IRAS100_NHI}, we see no correlation between the (filtered) pixel values of the reconstructed CIB/LSS map and the purely Galactic HI map, with a Pearson correlation coefficient of $0.005$. Our reconstructed CIB map is thus foreground-free to high fidelity. As the issue of foreground mitigation has been the main challenge in the whole emerging field of intensity mapping \citep[e.g.,][]{2017arXiv170909066K}, having an effective component separation method, as introduced in this work, could potentially be a game changer in observational cosmology. A foreground-free LSS map would not only enhance the significance of any cross-correlation measurements (e.g., Figure~\ref{fig:rec_LSS_x_SDSS}), but it will also enable robust LSS auto-correlation measurements that were not possible before.

\subsection{Is CSFD Truly LSS-Free?} \label{sec:CSFD}

The next question is: How clean is our CSFD dust map, and to what extent is it free from the extragalactic imprints originally seen in SFD? To address this, we first perform a sanity check by feeding the new CSFD dust map into the same tomographic cross-correlation pipeline (Sections~\ref{sec:extract_stats_SFD_1D} and \ref{sec:extract_stats_SFD_2D}) based on the SDSS galaxies and quasars as references. The result is shown in Figure~\ref{fig:Summary}-C, to be compared directly with that applied to the original SFD in Figure~\ref{fig:Summary}-B. We can see that the excess $E(B-V)$ in CSFD around the references is everywhere consistent with zero in both the 1D and 2D panels, which is in stark contrast with the significant contamination in SFD, although the two maps look visually indistinguishable, and the CSFD dust map still carries all the Galactic features (e.g., the strong latitude gradient and the cirrus). This confirms that our SFD correction based on the LSS reconstruction is doing exactly what we expect it to do, i.e., taking out all detectable LSS signatures as measured by the SDSS references up to at least $z=3$ and $\theta = 4^{\circ}$, and the CSFD is significantly cleaner than SFD. Our CIB cleaning, in fact, goes beyond these redshift and angular scales, as there is a natural extrapolation provided by our LSS templates. We note that the difference between panels B and C in Figure~\ref{fig:Summary} is exactly what is shown in Figure~\ref{fig:rec_LSS_x_SDSS} for the reconstructed CIB map.

\begin{figure*}[ht!]
    \begin{center}
         \includegraphics[width=0.78\textwidth]{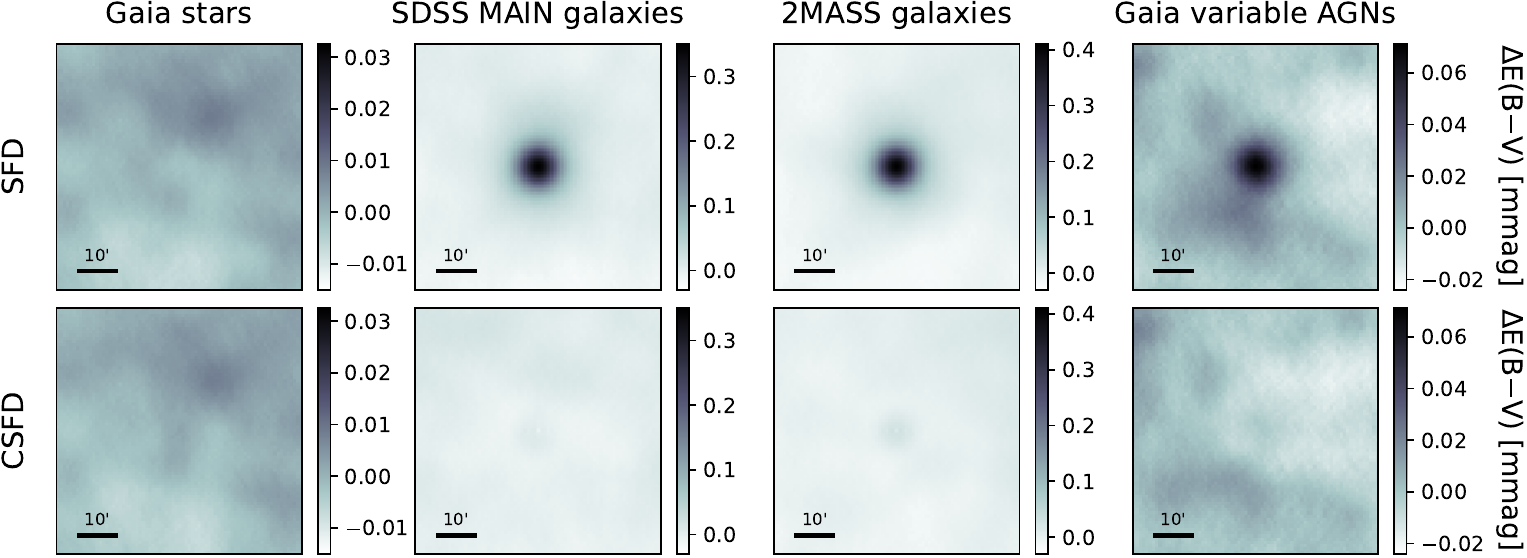}
    \end{center}
    \caption{Stacked images of the $E(B-V)$ in SFD in the upper row and CSFD in the lower row. From left to right, the stacks are centered on Gaia stars, SDSS MAIN sample galaxies, 2MASS galaxies, and Gaia variability selected AGNs, respectively. Except for the stars, the $E(B-V)$ in SFD around all these galaxy samples are biased high due to the LSS contamination. Such bias is significantly reduced in CSFD.}
    \label{fig:stacks-1}
\end{figure*}

\begin{figure*}[ht!]
    \begin{center}
         \includegraphics[width=0.78\textwidth]{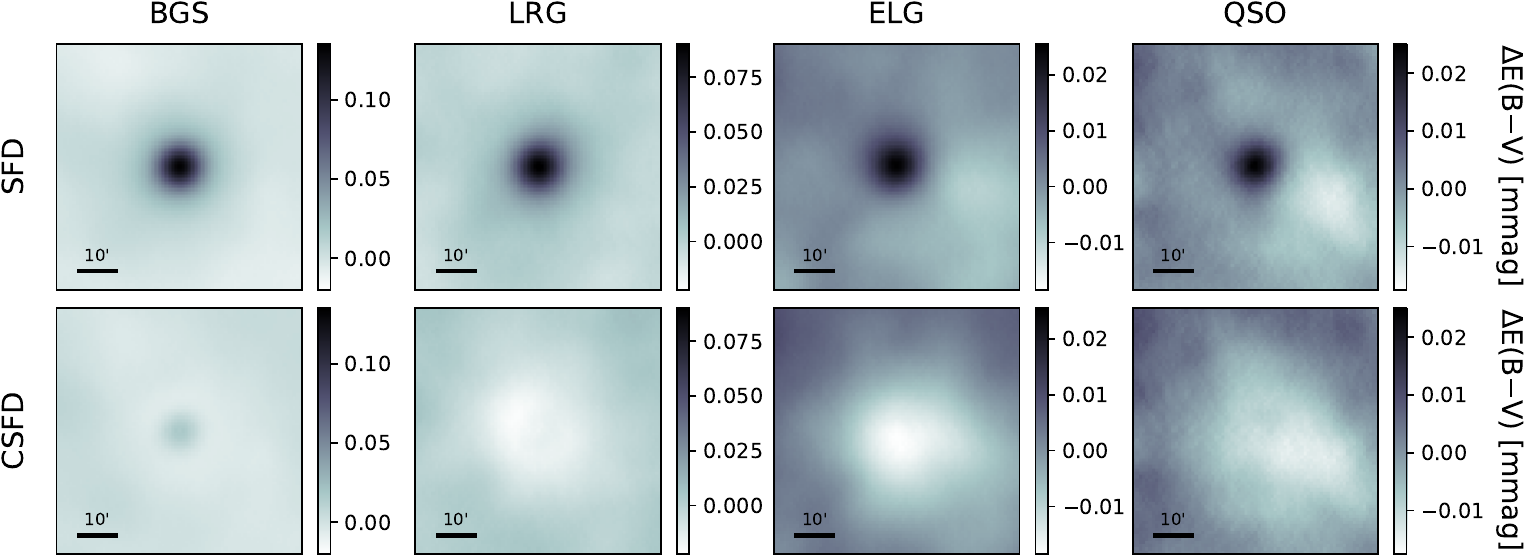}
    \end{center}
    \caption{Same as Figure~\ref{fig:stacks-1} but now centered on four samples of DESI targets, BGS, LRG, ELG, and QSO, from left to right columns, respectively. Similar to that in Figure~\ref{fig:stacks-1}, the CSFD dust map has a much lower residual CIB contamination than SFD. In addition, the CSFD stacks (lower row) reveal that LRG, ELG, and QSO are preferentially selected against high extinction patches of the sky as they push the faint magnitude limits of the DESI Legacy imaging surveys. Such a selection systematic can be removed by re-thresholding the DESI samples using CSFD.}
    \label{fig:stacks-DESI}
\end{figure*}

We now validate the LSS removal in CSFD via imaging stacking, which is itself a cross-correlation analysis but focuses more on visualizing small-scale behaviors. In Figure~\ref{fig:stacks-1}, we show stacks of the $E(B-V)$ values in both SFD (upper row) and CSFD (lower row) in the unmasked cosmology area for four different samples. These include, from left to right, Gaia stars \citep{2023A&A...674A...1G}, SDSS MAIN sample galaxies \citep{2002AJ....124.1810S, 2005AJ....129.2562B, 2015MNRAS.449..835R}, Two Micron All Sky Survey (2MASS) galaxies \citep[extended sources from][]{2006AJ....131.1163S}, and variability selected AGN candidates in Gaia DR3 \citep{2023A&A...674A..41G}. The images are $1^{\circ}$ on a side. In each panel, the map median outside the central $r=10'$ area is taken as the zero-point and subtracted to show the deviations. To achieve higher SNRs and be robust against foreground, for both SFD and CSFD, we apply the HI-based map-level cleaning described in Appendix~\ref{apd:filtering_HI_clean} and restrict the sky area to that with $|b|>60^{\circ}$. Among these four sets of stacks, Gaia stars show no correlation with SFD to start with, nor CSFD. This suggests that the Galactic dust in the ISM does not correlate with stars strongly on a 1$^{\circ}$ scale. For the remaining three extragalactic samples, we can see that they are significantly ``detected'' in SFD due to the LSS contamination. Nicely, the excess $E(B-V)$ is largely gone in CSFD, reassuring us that our LSS removal is effective. The SDSS MAIN sample galaxies are used in constructing CSFD as part of our reference sample, so their stacks are considered a sanity check. Both the 2MASS galaxies and Gaia variable AGNs are not directly used in constructing CSFD, neither in the reference nor the LSS templates; these stacks are, therefore, independent validations of the fidelity of our LSS cleaning.

There is still a small yet positive residual $\Delta E(B-V)$ in CSFD at the locations of 2MASS galaxies with a compact angular extent close to the beam. Nonetheless, the peak amplitude has been reduced to 7\% from that in the original SFD. The picture is likely as follows: although we have successfully removed most, if not all, of the two-halo term clustering of the CIB in SFD, a perfect removal of the one-halo term would require directly resolving the CIB into individual sources in the FIR at a flux limit and angular resolution far beyond IRAS (and AKARI), which is currently not possible over wide fields. In CSFD, the level of the one-halo term residual depends on the kind of sources being stacked and is generally limited by the fact that not all FIR bright galaxies each share their host halos with at least one NIR selected galaxy down to the flux limit of our WISE LSS sample. Despite this limitation, our new CSFD dust map is still much cleaner than SFD in the one-halo regime and is free from the extended LSS on superhalo scales.


\section{Discussion} \label{sec:discuss}

\subsection{Example Application for DESI Targets} \label{sec:CSFD-DESI}
As this paper is motivated by the need to improve Galactic extinction estimates for cosmology applications, as an example, here we investigate the cosmology targets in the ongoing DESI survey. In Figure~\ref{fig:stacks-DESI}, we show the SFD versus CSFD stacks (at $|b|>60^{\circ}$), the same as in Figure~\ref{fig:stacks-1}, now for four samples of photometric targets in DESI. As labeled in the plot, from left to right, these are BGS \citep{2023AJ....165..253H}, LRG \citep{2023AJ....165...58Z}, ELG \citep{2023AJ....165..126R}, and QSO \citep{2023ApJ...944..107C}, respectively. These targets are selected using the DR9 of the DESI Legacy Imaging Surveys \citep{2019AJ....157..168D} prior to DESI's nominal spectroscopic observations. We can see that all four DESI target samples are strongly ``detected'' in SFD due to its LSS contamination. In the corresponding CSFD stacks shown on the lower row, the LSS imprints are largely removed. This is, again, reassuring that our CIB removal is effective. Similarly, there are some compact, beam-scale residuals at the centers of the images of low levels, especially for BGS (13\% peak amplitude compared to the SFD stack), likely due to the imperfect removal of the one-halo term CIB. 

Interestingly, we notice another weak but significant feature. At scales more extended than the beam but smaller than a degree, we see negative $\Delta E(B-V)$ around LRG, ELG, and QSO in the CSFD stacks. This reveals a selection effect that these samples are preferentially at slightly low dust column patches of the sky (or against high extinction areas). In fact, such a selection effect is not surprising given that these targets, especially the ELGs, are pushing the faint magnitude limits of the imaging survey, and the completeness is thus low in high-dust column areas. This effect is discussed as a potential concern in the DESI ELG selection in \cite{2023AJ....165..126R} and has perhaps already introduced systematics in the ELG component of SDSS-IV eBOSS as discussed in \cite{2017MNRAS.465.1831D}. Interestingly, such a local underdensity of dust around faint targets is not seen in the SFD stacks in the top row in Figure~\ref{fig:stacks-DESI}; it appears that the positive CIB contamination from the extended two-halo term happens to fill in the deficit. All in all, since selecting against high extinction regions is a real systematic effect that needs to be mitigated, we suggest simply using our new CSFD dust map to re-threshold the samples (basically redo the extinction correction) after the spectroscopic redshifts are obtained in DESI but before further cosmology inference. Alternatively, this selection effect can be captured in the construction of the random catalogs if CSFD is used for the forward modeling simulations. The DESI collaboration has agreed to replace the original SFD with our new CSFD map for their downstream sample selection and cosmology analyses (DESI collaboration, private communication).

\subsection{Future Improvements}
\label{sec:future_work}

Although the CSFD dust map is a significant advancement upon SFD, there are several aspects that can be improved in the future. Here we discuss prospects for the angular resolution, large-scale calibration, and further iterations in the LSS removal.

The $6.1'$ beam of SFD/CSFD generally adds noise to extinction corrections in optical astronomy where the resolution is usually of the order of an arcsecond. To improve, we discuss two avenues. First, we can consider filling in the small-scale, i.e., high-$k$ power from the 15" resolution diffuse WISE 12~$\mu$m (W3) map from \cite{2014ApJ...781....5M}, which traces a strong feature of the polycyclic aromatic hydrocarbons (PAHs) in the Galactic ISM. Since W3 is broadband, the merged CSFD-W3 map would also contain significant extragalactic imprints just like that in SFD (see CM19 for the detection of the cosmic PAH background). Another pass of LSS cleaning using the tomographic method introduced in this paper is thus needed. A second promising way is to use the data from the upcoming SPHEREx mission \citep{2018SPIE10698E..1UK}, which covers a 3.3~$\mu$m PAH feature over the full sky with a $6''$ resolution using spectrophotometry with a spectral resolving power $R=35$. This is sufficient to resolve Galactic PAHs from the extragalactic background spectrally, so no further LSS cleaning is needed. No matter which PAH map is used for the inpainting, one should still use CSFD at scales beyond $6.1'$ as PAHs, although they correlate with dust well (much better than HI), are still not direct tracers of the total dust column.  

By construction, CSFD conserves all properties of SFD beyond 20$^{\circ}$, as our LSS templates are filtered at that scale (Appendix~\ref{sec:templates}). As mentioned in Section~\ref{sec:intro}, a $\sim 14\%$ global zero-point offset in $E(B-V)$ has been suggested by \cite{2011ApJ...737..103S} based on calibration using stellar spectra in SDSS. The same authors have also pointed out a North--South asymmetry, while the statistics are insufficient to derive a map-level correction on SFD. Independently, \cite{2010ApJ...719..415P} used the colors of red galaxies in SDSS as ``standard crayons'' to calibrate SFD and derived a large-scale correction map over the northern SDSS footprint at 4.5$^{\circ}$ resolution. They showed that the SFD reddening scale is overall spatially stable but can deviate from their calibration in some regions by up to a few tens of millimagnitudes in $E(B-V)$ depending on the dust temperature. These calibration methods are established and effective. The critical next step is then to apply them using new datasets, e.g., Gaia, the upcoming SPHEREx, or other surveys to perform full-sky calibrations on the spatially varying reddening zero-point of SFD/CSFD.

In our LSS removal for SFD, one aspect that can be improved is a slight inhomogeneity in the selection of our reference sample (Appendix~\ref{sec:ref}). When each of these SDSS spectroscopic LSS catalogs was constructed, extinction correction was already applied, which was done using SFD for all cases. The threshold color-magnitude cuts in the selection are then affected by the spatial- and redshift-dependent bias in SFD due to the CIB contamination, the very effect that we are fixing in this paper. As a result, our tomographic cross-correlations could be slightly biased. However, the amplitude of this effect should be small, and the impact could probably be absorbed into modulating the galaxy bias, including its scale dependence. Since the same happens when cross-correlating the references with both SFD and the LSS templates, the effect should be largely canceled. We thus do not intend to treat or mitigate it in this work. With that said, improving the selection homogeneity of the references (and also the LSS templates) is straightforward with an iterative approach. We can re-threshold the SDSS references based on photometry dereddened using the newly constructed CSFD dust map and perform another pass of the LSS cleaning using exactly the same procedure in this paper. Since DESI will provide a much larger spectroscopic reference sample soon and they will use CSFD to ensure an accurate extinction correction, one could revisit the LSS removal again in the future. With the references from DESI and LSS templates using deeper WISE coadds or, e.g., SPHEREx, we could expect a higher SNR CIB reconstruction and cleaning.

\subsection{CSFD as a New Standard Dust Map}
\label{sec:sfd_to_csfd}

Since 1998, SFD has been the default dust map for Galactic extinction correction in the UV, optical, and NIR for both large cosmological experiments and individual users. In recent years, more than a dozen new dust maps came online thanks to new data from $Planck$, $Gaia$, HI experiments, and more. These new maps greatly improve our knowledge about different aspects of the Galactic ISM. Yet, none are routinely used for extinction correction, and SFD remains the go-to map 25 yr after its release. In addition to following the convention and for consistency, we see the following reasons for the use of SFD. First, compared to most $Planck$-based FIR emission dust maps \citep[e.g.,][]{2015ApJ...798...88M,2016A&A...586A.132P}, SFD has less CIB contamination as shown in CM19. Compared to stellar reddening maps from Pan-STARRS and Gaia \citep[e.g.,][]{2019ApJ...887...93G,2023ApJ...949...47M}, SFD is of far higher SNR, which leads to less noise in extinction corrections. Compared to full-sky HI-based reddening maps \citep[e.g.,][]{2017ApJ...846...38L}, SFD has a higher angular resolution and traces dust column more directly. 

In this work, we design our CSFD---the corrected SFD dust map---to carry all of the advantageous properties of SFD while further improving upon the main issue in extragalactic contamination. That is to say, CSFD is in almost every aspect the same as the original SFD but with significantly suppressed CIB. Therefore, the new CSFD map is in a better position for the community to lower the risk of unwanted systematics in upcoming cosmology analyses with galaxy surveys, supernovae Ia, CMB experiments, as well as intensity mapping. Large galaxy evolution studies, especially in the UV and optical, would benefit from more accurate extinction correction provided by CSFD with no LSS-correlated overcorrection. For many other users who study a small number of targeted objects, there could be no noticeable difference in using SFD versus CSFD. All in all, implementing the switch from SFD to CSFD is only beneficial and straightforward; since the map unit, the calibration, and most of the map properties are the same, one can directly swap the dust maps with minimal, if not zero change needed in existing and future data analysis pipelines.

Another way that CSFD could be valuable is to serve as a clean Milky Way foreground template for CMB experiments, e.g., CMB-S4 \citep{2019arXiv190704473A} and intensity mapping of the extragalactic background light (EBL) with, e.g., SPHEREx \citep{2014arXiv1412.4872D} and the Square Kilometre Array \citep{2009IEEEP..97.1482D}. In almost all wavelengths, the foreground is dominated by dust or other components in the ISM that correlate with dust. A map-level regression or power spectrum modeling using CSFD as the foreground template thus provides the most effective mitigation, with higher precision than HI and better accuracy than the original SFD or the $IRAS$ channel maps as foreground templates.

\subsection{Data Release}
\label{sec:data_release}
We describe the data package on our project website,\footnote{\url{https://idv.sinica.edu.tw/ykchiang/CSFD.html}} and release it on Zenodo.\footnote{\url{https://zenodo.org/record/8207175}} A copy is available on NASA LAMBDA.\footnote{\url{https://lambda.gsfc.nasa.gov/product/foreground/fg_csfd_reddening_map_info.html}} Map value queries can be done using the \texttt{dustmaps} package \citep{2018JOSS....3..695G}.\footnote{\url{https://dustmaps.readthedocs.io/}} The 
data include the following maps, all in HEALPix $N_{side}=2048$ resolution with ring ordering in the Galactic coordinates:
\begin{enumerate}
    \item CSFD, the  LSS-free full-sky Galactic dust map in $E(B-V)$. The reddening scale is the same as the original SFD with no renormalization.
    
    \item The original SFD recast in the HEALPix format.
    
    \item Reconstructed LSS or the foreground-free 100~$\mu$m CIB intensity map in $\rm MJy\ sr^{-1}$. There is no monopole by construction. The beam is consistently $6.1'$ (including the small fraction of the sky where SFD and CSFD are of DIRBE resolution).
    
    \item Error map of the reconstructed LSS/CIB field.
    
    \item Bitmask indicating the LSS correction footprint, the most reliable cosmology area, and the lower-resolution, no IRAS area.
\end{enumerate}

\section{Conclusions} \label{sec:conclusion}

Accurate mapping of the Galactic dust reddening field over the entire sky is an important community task, as extinction correction is needed for essentially all extragalactic photometry in the UV, optical, and NIR. With more ambitious sky surveys coming online, the accuracy of the widely used \citet*[][SFD]{1998ApJ...500..525S} dust map needs to be revisited. One major issue discovered and rediscovered in the literature is the CIB contamination in SFD due to emission from unresolved dusty star-forming galaxies \citep{2007PASJ...59..205Y,2019ApJ...870..120C}, and the imprint is two-halo-term dominated at the SFD resolution. When SFD is used for extinction correction, the CIB residuals introduce scale- and redshift-dependent overcorrection, which can bias galaxy clustering, lensing, supernova Ia distances, and the downstream cosmology inference. In this work, we build on the foundation of SFD and focus on deriving a CIB-free dust map.

To clean SFD, we carry out a map-level tomographic reconstruction of the CIB (at least the two-halo contribution) and remove it from the reddening estimates. The end product is a more accurate full-sky Milky Way dust map dubbed CSFD---the corrected SFD. Our analysis consists of the following steps:
\begin{enumerate}
    \item First, we get summary statistics of the CIB in SFD by cross-correlating SFD's $E(B-V)$ field with 2.7 million references using spectroscopic galaxies and quasars in  SDSS as functions of redshift and scale.
    
    \item For a map-level CIB reconstruction, information on the Fourier phases of the cosmic web is needed, which we obtain from 180 full-sky LSS templates,  a basis set built using over 600 million WISE galaxies over $z=0$--3 sliced in observed properties and augmented with a range of beam sizes. A supplemental template is added to account for FIR bright galaxies detected in AKARI at $z<0.1$.

    \item We derive a reconstructed CIB map using a linear combination of these $181$ LSS templates such that the tomographic two-point statistics are indistinguishable from that of the CIB in SFD. 

    \item Finally, we subtract the reconstructed CIB map out of SFD, which leads to the new, LSS-cleaned CSFD dust map.
\end{enumerate}

The correction term, i.e., the reconstructed CIB map, is itself a valuable product for tracing the matter density field in the Universe. We achieve a mean $\rm SNR \sim 3$ for every spatial resolution element and a combined SNR of the order of thousands. Via cross-correlating with a Galactic HI 21~cm map, we show that our reconstructed CIB map is indeed foreground-free to high fidelity. This is in stark contrast with, e.g., the $IRAS$ 100~$\mu$m channel map and most other EBL maps where the foreground  has been hampering cosmological applications.

We validate CSFD, our new Galactic dust reddening map, via stacking various star, galaxy, AGN, and quasar samples and show that the imprints of LSS are largely removed. Although there are small one-halo term residuals (for bright, low-redshift galaxies), the CSFD map is significantly cleaner and more faithfully traces the true Galactic reddening field than the original SFD. As an example application in cosmology experiments, we show that CSFD reveals a selection effect in DESI, where faint, high-redshift targets are preferentially selected in slightly low extinction regions. Such systematics can now be readily removed by re-applying extinction corrections using CSFD before further cosmology analyses in DESI. Since CSFD has minimal extragalactic contamination, it can also serve as a new template to mitigate the Galactic foreground in CMB and intensity mapping experiments.

The SFD dust map is likely sufficient for most galaxy evolution studies at high latitudes but not precision cosmology. The new CSFD Galactic dust reddening map carries most of the properties of the original SFD, including the map unit and calibration, sky coverage, SNR, and large-scale zero-points. On small scales, CSFD is significantly more accurate; it minimizes the risk of potential CIB-induced biases in downstream analyses. For these reasons, we conclude that replacing SFD with CSFD for future Galactic extinction corrections is both straightforward and beneficial for more accurate cosmology and astrophysics.

\begin{acknowledgments}
Y.C. acknowledges the support of the National Science and Technology Council of Taiwan through grant NSTC 111-2112-M-001-090-MY3. The author thanks Brice M\'enard and David Schlegel for helpful discussions, the anonymous referee for valuable comments, Rongpu Zhou for providing the DESI target catalogs, Gregory Green for incorporating CSFD into the \texttt{dustmaps} package, and Graeme Addison for uploading CSFD to the NASA LAMBDA archive. 

\end{acknowledgments}

%






\appendix

\section{Data Preparation} \label{sec:data}
Here we describe the data preparation for the SFD map, the cross-correlation reference, and the LSS templates used in Section~\ref{sec:comp_sep_SFD}.

\subsection{SFD} \label{sec:data-SFD}

For the SFD map used in this work, we recast the $E(B-V)$ field from \cite{1998ApJ...500..525S} originally provided in the Lambert projection to a full sky HEALPix \citep{2005ApJ...622..759G} map with $N_{side}=2048$ resolution, which is visualized in Figure~\ref{fig:Summary}-A (lower-left). The dynamic range of the $E(B-V)$ values is large ($10^5$ in magnitude) across Galactic latitudes, while at $|b|>50^{\circ}$, the mean $E(B-V)$ is about 20~mmag.

The factor of 0.86 (14\%) re-scaling mentioned in Section~\ref{sec:intro} derived by \cite{2011ApJ...737..103S} is not applied, as many existing survey photometry pipelines already include this factor downstream by post-processing, and it would reduce the friction for users if we design our new CSFD dust map to carry the calibration of the original SFD. 


The SFD map is used as $I(\phi)$ in two places in our formalism in Section~\ref{sec:formalism}, firstly in measuring the tomographic LSS statistics (Equation~\ref{eq:5}), and secondly in the final component separation (Equation~\ref{eq:9}). In practice, we can apply some foreground mitigation or cleaning in the former case to increase the SNRs and reduce the foreground-induced bias. The mitigation scheme can be very aggressive as long as it does not alter the LSS component ($I_{LSS}(\phi)$ in Equation~\ref{eq:1}). We apply an HI-based map-level foreground cleaning to SFD, which will be described in Appendix~\ref{apd:filtering_HI_clean} together with other cross-correlation optimizations. For the latter case in the final reconstruction (Equation~\ref{eq:9}), the original, unprocessed SFD is used directly as $I(\phi)$.

\subsection{SDSS References} \label{sec:ref}

To characterize the scale- and redshift-dependent CIB in SFD using cross-correlations, here we define our reference sample ($R$ in Section~\ref{sec:formalism-stats}). Since the purpose is to extract statistics, the cosmological principle applies; even though we aim for a full-sky CIB reconstruction, the reference $R$ does not need to be full-sky.

\begin{figure}[t!]
    \begin{center}
         \includegraphics[width=0.45\textwidth]{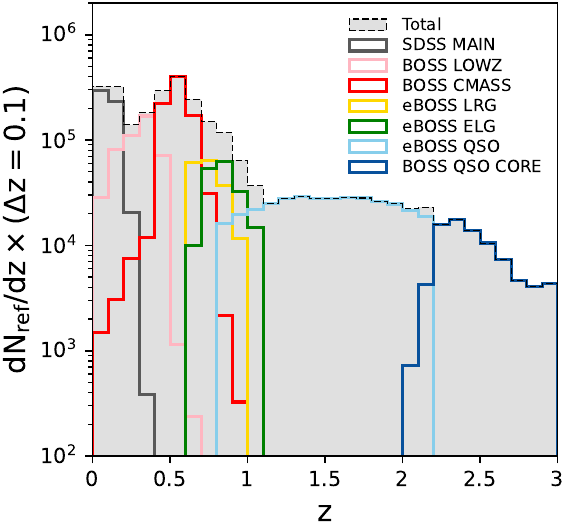}
    \end{center}
    \caption{Redshift distributions for the cross-correlation reference sample and subsamples from spectroscopic galaxies and quasars in SDSS.}
    \label{fig:ref}
\end{figure}

We compile a set of reference sources with spectroscopic redshifts using the ``LSS samples,'' i.e., clustering analysis ready samples, from the original SDSS, the Baryon Oscillation Spectroscopic Survey (BOSS), and the Extended Baryon Oscillation Spectroscopic Survey (eBOSS). These are split into seven subsamples ordered by increasing redshift as shown in Figure~\ref{fig:ref}: SDSS MAIN \citep{2002AJ....124.1810S, 2005AJ....129.2562B, 2015MNRAS.449..835R}, BOSS LOWZ and CMASS \citep{2016MNRAS.455.1553R}, eBOSS LRG, ELG, and QSO \citep{2020MNRAS.498.2354R,2021MNRAS.500.3254R}, and BOSS QSO CORE \citep{2012MNRAS.424..933W, 2015MNRAS.453.2779E}. For the spatial selection function or masking information, we collect the associated random catalogs when available. For those where the randoms are not directly provided, we reconstruct them following the procedure laid out in these references listed for each catalog. Data from both the Northern and Southern SDSS fields are included. Although the detailed footprints vary among subsamples, the selection functions captured in the random catalogs can be added linearly. We, therefore, combine all the subsamples into one effective reference set, totaling 2.7 million spectroscopic objects in $0 < z < 3$ over $\sim10,000$ deg$^2$, a quarter of the sky.

\subsection{WISE LSS Templates} \label{sec:templates}

As introduced in the formalism in Section~\ref{sec:formalism-recon}, the LSS template set $T_{LSS, i}(\phi)$ provides the phase information for the pattern of the LSS in the sky. Here we describe practical considerations in the case of reconstructing the CIB in SFD. We have several requirements for $T_{LSS, i}(\phi)$:

\begin{enumerate}
    \item The template set should be of high SNR, foreground-free, and (nearly) full-sky.
    \item The redshift ranges of individual templates should be diverse, while together, they should cover the full span of the CIB in SFD from $z=0$ to 3.
    \item The templates' small-scale power spectra should span a range of shapes such that the combination can mimic the nonlinear clustering of the CIB.
\end{enumerate}
Requirement 1 ensures our end-product CSFD dust map is of high precision and maximally applicable to extinction corrections for surveys of any footprint. For $T_{LSS, i}(\phi)$ to be foreground-free, we consider building them using only individually detected sources but not diffuse sky intensity maps. Requirements 2 and 3 elaborate on the key idea that the templates need to form a complete basis set for reconstructing the 100~$\rm \mu m$ CIB in SFD. The basis set can be overcomplete and does not need to be over-engineered, as we do not require the templates to be linearly independent. The strategy is thus to increase diversity without removing redundancy.

We consider a two-step approach to building the LSS templates $T_{LSS, i}(\phi)$. First, to meet requirements 1 and 2, we construct a set of 30 so-called ``base templates'' using photometric galaxy density fields in the WISE mission \citep{2010AJ....140.1868W} beam-matched to SFD. For requirement 3, we then make six ``augmented templates'' per base with different amounts of beam-smearing. The total number of WISE LSS templates is thus $30\times 6= 180$.

\subsubsection{Base from Two WISE Catalogs} 
\label{apd:base_templates}

The task to sample the cosmic web densely over the full sky up to $z=3$ might sound daunting, while at the stage of building the templates, we only need minimal information that a large set of high-redshift objects exist and can be selected in the catalog. Data in the WISE mission are thus suitable and meet essentially all requirements. We use two complementary WISE catalogs: (1) CatWISE2020 \citep{2021ApJS..253....8M} to access the largest number of sources to suppress the shot noise and to go up to the highest possible redshifts, and (2) the WISE$\times$SuperCOSMOS photometric-redshift catalog \citep{2016ApJS..225....5B} to sample the low-redshift Universe with a finer redshift resolution. We note that, at this stage, we do not need to know the exact redshifts of these sources, as the tomographic cross-correlations we apply later will provide such information, which will then be incorporated implicitly in the reconstruction.

\begin{figure}[t!]
    \begin{center}
         \includegraphics[width=0.45\textwidth]{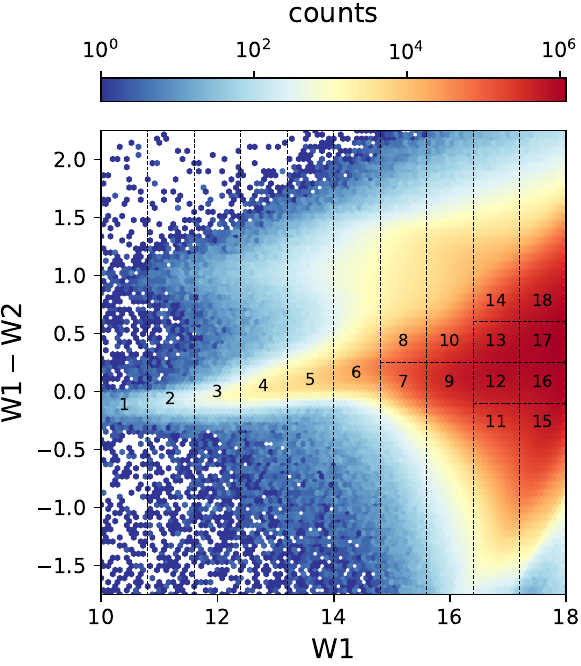}
    \end{center}
    \caption{Galaxy number density in the $W1$ versus $W1-W2$ color-magnitude space of the full-sky CatWISE2020 catalog \citep{2021ApJS..253....8M} in our LSS reconstruction footprint. Dashed lines define the 18 cells we use to build our first part of the base LSS templates.}
    \label{fig:CatWISE}
\end{figure}

Although WISE data are full-sky, in addition to the Galactic plane and Magellanic clouds, there are regions where the extragalactic information is unreliable. These include areas near bright stars and large galaxies, as well as those affected by bright dust cirrus and other WISE-specific imaging artifacts. We build a mask by merging a generic WISE extragalactic mask \citep[][with some small, single-pixel bad areas ignored]{2020JCAP...05..047K}, a threshold stellar density mask based on Gaia DR3 \citep{2023A&A...674A...1G}, and the mask associated with the  WISE$\times$SuperCOSMOS catalog \citep{2016ApJS..225....5B}. This leads to the two tiers of cosmology and noncosmology areas within the footprint of our reconstructed CIB map in Figure~\ref{fig:Summary}-A (upper), where the latter is masked as a transparent field.

The CatWISE2020 catalog is the deepest WISE-based catalog currently available. It contains nearly 2 billion sources in two bands, 3.4~$\rm \mu m$ (W1) and 4.6~$\rm \mu m$ (W2), with data taken from 2010--2018. We select all sources with $10 < W1 < 18$ and remove stars using data from Gaia DR3 \citep{2023A&A...674A...1G}\footnote{All Gaia objects not in the \textit{qso\_candidates} and \textit{galaxy\_candidates} tables are considered as stars.}. To avoid overcorrection of the LSS in SFD due to double counting, for bright sources with $W1<14.8$, we remove those in the FIR point-source regions already masked by \cite{1998ApJ...500..525S}. For all WISE sources in our sample, we also remove those matched (within 20$''$) to a reliable subset of AKARI galaxies that will be described in Appendix~\ref{sec:akari}. This results in 584 million CatWISE galaxies or AGNs in the footprint of our LSS reconstruction. To form an ideal basis set of the CIB in SFD, we split these sources into subsamples of 18 cells in the color-magnitude diagram (CMD) as visualized in Figure~\ref{fig:CatWISE}. The exact divisions of these color-magnitude cells are not critical, as the goal is just to obtain diverse populations in terms of redshift, galaxy bias, and halo occupation, which naturally come out if we split the sample in photometric properties. The sizes of the cells are also somewhat arbitrary, but they are no smaller than the typical photometric errors at the faint end.

As demonstrated in CM19, the amplitude of the CIB in SFD increases rapidly toward lower redshift. The LSS basis set should thus be as flexible as possible at low redshift, with templates each spanning a narrow redshift range. We, therefore, supplement the 18 CMD splits of CatWISE objects with galaxies in 12 photometric-redshift (photo-z) bins from the WISE$\times$SuperCOSMOS catalog. This catalog is constructed using earlier, and thus shallower WISE photometry \citep[AllWISE;][]{2013wise.rept....1C} but combined with full-sky data from SuperCOSMOS \citep{2016MNRAS.462.2085P} in the optical and 2MASS \citep{2006AJ....131.1163S} in the NIR to constrain redshifts. The same as for the CatWISE case, we remove those in the SFD point-source mask and the ones matched to AKARI galaxies. This results in a total of 20 million WISE$\times$SuperCOSMOS objects up to photo-z of 0.5 in our LSS correction footprint. We set the photo-z bin edges at [0, 0.02, 0.04, 0.06, 0.08, 0.1, 0.15, 0.2, 0.25, 0.3, 0.35, 0.4, 0.5], while again, the exact binning scheme is not critical. We note that the photo-z here is merely a feature to split galaxies into diverse populations, just like colors and magnitudes; the robustness of our CIB reconstruction would not depend on these photo-z's being unbiased, as the redshift information will be obtained when cross-correlating with the SDSS references with spectroscopic redshifts.

\begin{figure*}[t!]
    \begin{center}
         \includegraphics[width=0.95\textwidth]{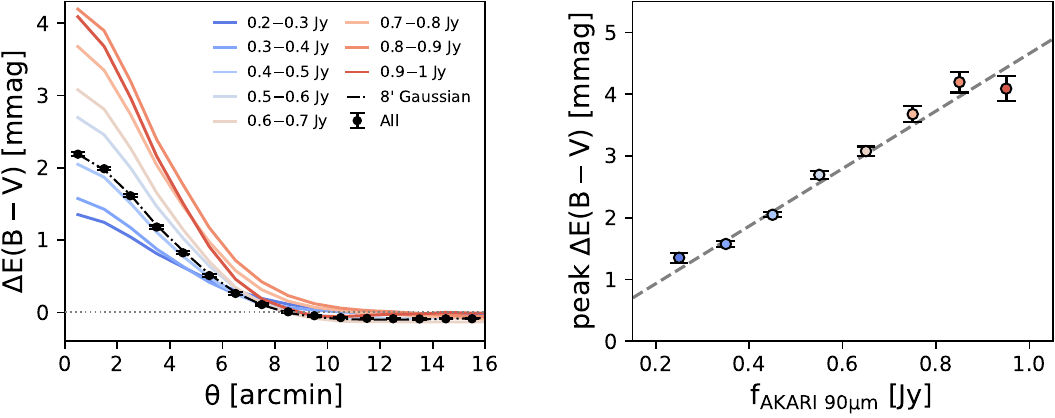}
    \end{center}
    \caption{Stacked SFD $E(B-V)$ profiles (left) and peak values (right) at the locations of secured AKARI galaxies in our cosmology area binned in 90~$\mu$m flux. The profiles are consistent with an $8'$ Gaussian, slightly more extended than the SFD beam, and the peak excess $E(B-V)$ scales linearly with 90~$\mu$m flux. This forms the basis for constructing our zeroth LSS template.}
    \label{fig:AKARI}
\end{figure*}

With these over 600 million galaxies in the $18+12$ CatWISE and WISE$\times$SuperCOSMOS subsamples, we create 30 base template maps $T_{LSS, i}^{base}(\phi)$ using the following steps. First, for each template, a map of source number density contrast is created using the HEALPix scheme with $N_{side}=2048$. We then smooth the map to match the $6.1'$ beam of SFD. To avoid large-scale systematics, we also apply a high-pass filter to remove fluctuations above 20$^{\circ}$ FWHM, which zeros out the large-scale mean of the maps. This filtering is done separately in the cosmology and noncosmology areas such that systematics in the latter would not leak into the former. Finally, we re-normalize the map values to have unit variance within the cosmology area, i.e., $Var(T)=1$ where $T$ is the dimensionless pixel values or ``temperatures'' of each template. This renormalization step is not critical, but it sets a convenient scale for $C_i$, the linear coefficients in Equation~\ref{eq:6} such that during the stage of the LSS reconstruction, $C_i$ can be directly interpreted as the relative amount of variance explained by each template $i$. We expect these 30 base templates $T_{LSS, i}^{base}(\phi)$ to not be entirely linearly independent, and have some redundancy. For example, the low-redshift Universe is covered by a subset of templates from both the CatWISE CMD cells and WISE$\times$SuperCOSMOS photo-z bins. The latter, however, is from brighter, and thus more clustered, galaxies in thinner redshift slices, which contribute differently to the 100~$\rm \mu m$ CIB in SFD than the former.

\subsubsection{Beam Augmentation} \label{sec:augmentation}

The halo occupation, and therefore small-scale clustering, of the CIB in SFD is effectively that for the global galaxy populations in the Universe weighted by their $100~\mu$m fluxes. This would generally differ from the halo occupations of the WISE galaxies we use in the templates. To be able to reproduce the small-scale behavior of the CIB, we take an empirical approach. We augment the 30 base templates each into six effective beam sizes: $1/\sqrt{2}$, $1$, $\sqrt{2}$, $2$, $2\sqrt{2}$, and $4$ times the $6.1'$ SFD beam with factors of $\sqrt{2}$ increments. For a given WISE subsample, the six augmented templates (including the base that is beam-matched to SFD) form a base set. This augmentation scheme effectively expands the basis space by making different high-$k$ cutoffs in the power spectra, thereby providing the flexibility to reconstruct the CIB with redshift-dependent power spectra of any shape. 


\subsection{AKARI FIR Bright Sources} \label{sec:akari}

We add an additional LSS template ($i=0$) to account for FIR bright galaxies previously unmasked in SFD. The full-sky AKARI mission \citep{2007PASJ...59S.369M} released 0.8 million sources detected at 90~$\mu$m as part of the Far Infrared Surveyor bright source catalog Version 2 \citep{2018cwla.conf..227Y}. As the source density correlates with Galactic structures, it is clear that some of these detections, especially the low-latitude ones, are spurious due to confusion with Galactic cirrus. We select a secure subset of 90~$\mu$m detected galaxies by cross-matching AKARI with extended sources in 2MASS \citep{2006AJ....131.1163S} and nonstellar (i.e., non-Gaia) WISE galaxies using a $20''$ matching radius. After removing those already masked in SFD, we obtain 38,266 AKARI galaxies within our LSS reconstruction footprint.

To build the point-source template explicitly using the flux information in AKARI, we stack the $E(B-V)$ field in SFD at the locations of these AKARI galaxies in our cosmology area.  To suppress the foreground, we aggressively filter SFD over 0.5$^{\circ}$ FWHM. Figure~\ref{fig:AKARI} shows the mean radial profiles and the peak excess $E(B-V)$ in bins of 90~$\mu$m flux in the left and right panels, respectively. We found that the stacked profiles are consistent with an $8'$ Gaussian (after the same 0.5$^{\circ}$ filtering), slightly more extended than the SFD beam. The peak amplitude in excess $E(B-V)$ scales linearly with the 90~$\mu$m flux in AKARI. Using this best-fit flux to $E(B-V)$ relation (dashed line in the right panel) and the $8'$ Gaussian profile, we then create a HEALPix map of the excess $E(B-V)$ due to these bright IR galaxies as our zeroth LSS template $T_{LSS,0}$. As the sky uniformity of AKARI detections is not guaranteed, and also to avoid changing the zero-point of the SFD cleaning, we remove the map mean over a 5$^{\circ}$ FWHM. For consistency with the WISE-based templates, we further normalize this AKARI map to have unit variance in our cosmology area, while during the LSS reconstruction (Section~\ref{sec:reconstruction}), $C_0$ for this AKARI point-source template is fixed to the exact factor that zeros out the stack excess $E(B-V)$ in Figure~\ref{fig:AKARI}.

\begin{figure*}[t!]
    \begin{center}
         \includegraphics[width=0.725\textwidth]{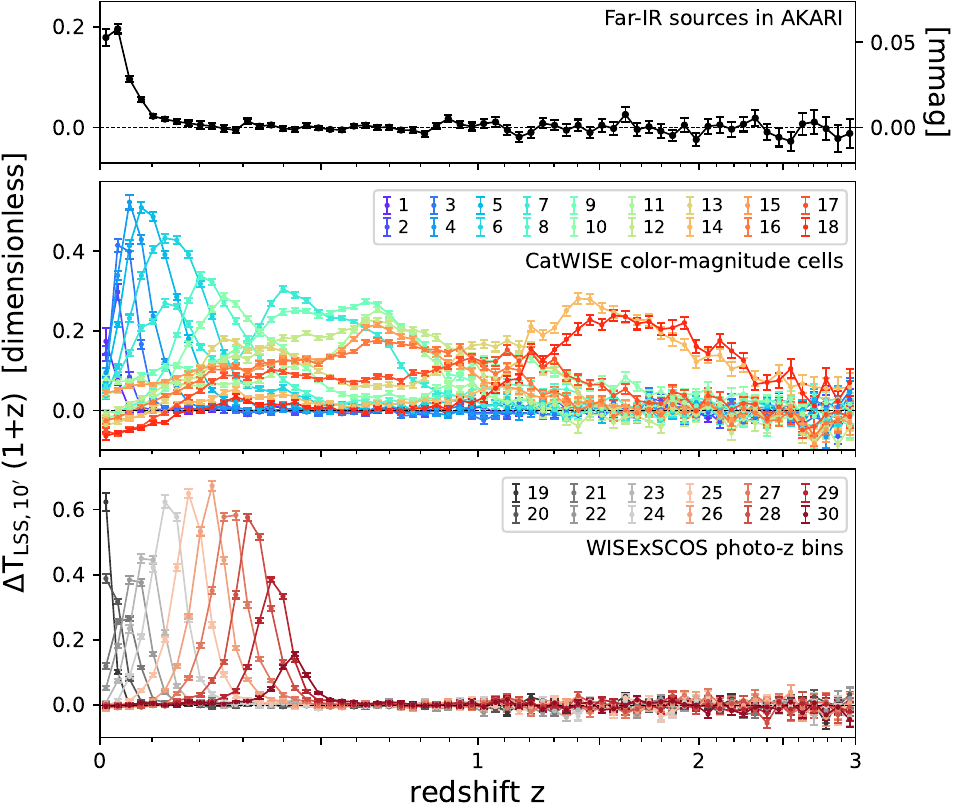}
    \end{center}
    \caption{Redshift tomography for the 30 LSS base templates $T_{LSS}(\phi)$ constructed using the 18 CatWISE samples (middle) and the 12 WISE$\times$SuperCOSMOS ones (bottom); that for the additional AKARI point-source template is shown in the top panel. The estimator extracts the excess dimensionless map values within $10'$ around the SDSS references, the same as that applied to SFD/CSFD in Figure~\ref{fig:Summary}, with an additional $1+z$ factor to better visualize the high-redshift part. The CatWISE templates are essential in covering the distant Universe, and the SuperCOSMOS ones offer finer sampling in the more local Universe, providing the basis we need to reconstruct the CIB in SFD. The AKARI point sources are restricted to mostly $z<0.1$.
     }
    \label{fig:LSSxRef1D}
\end{figure*}

\begin{figure*}[h!]
    \begin{center}
         \includegraphics[width=0.7\textwidth]{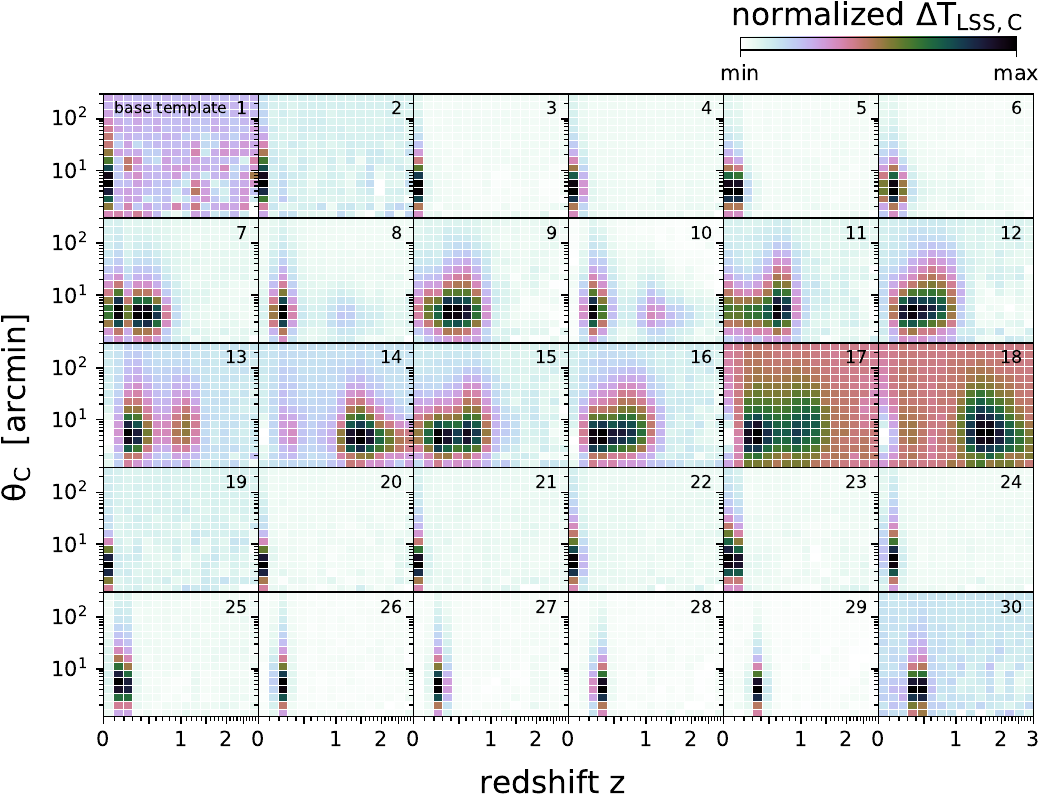}
    \end{center}
    \caption{2D angular and redshift tomography for the 30 base LSS templates $T_{LSS}(\phi)$ with the ID labeled in the upper right. Each pixel in this plot is a 2D cross-correlation with the SDSS references using the compensated filter, the same as in the lower B and C panels in Figure~\ref{fig:Summary} for SFD/CSFD. This plot shows that the templates contain clustered LSS over a wide range of redshifts and angular scales needed for the reconstruction. For a clear visualization, the color is normalized separately in each panel over the minimum and maximum cross-amplitudes $\Delta T_{LSS,C}(z, \theta_C)$.}
    \label{fig:LSSxRef2D}
\end{figure*}

\begin{figure*}[h!]
    \begin{center}
         \includegraphics[width=0.7\textwidth]{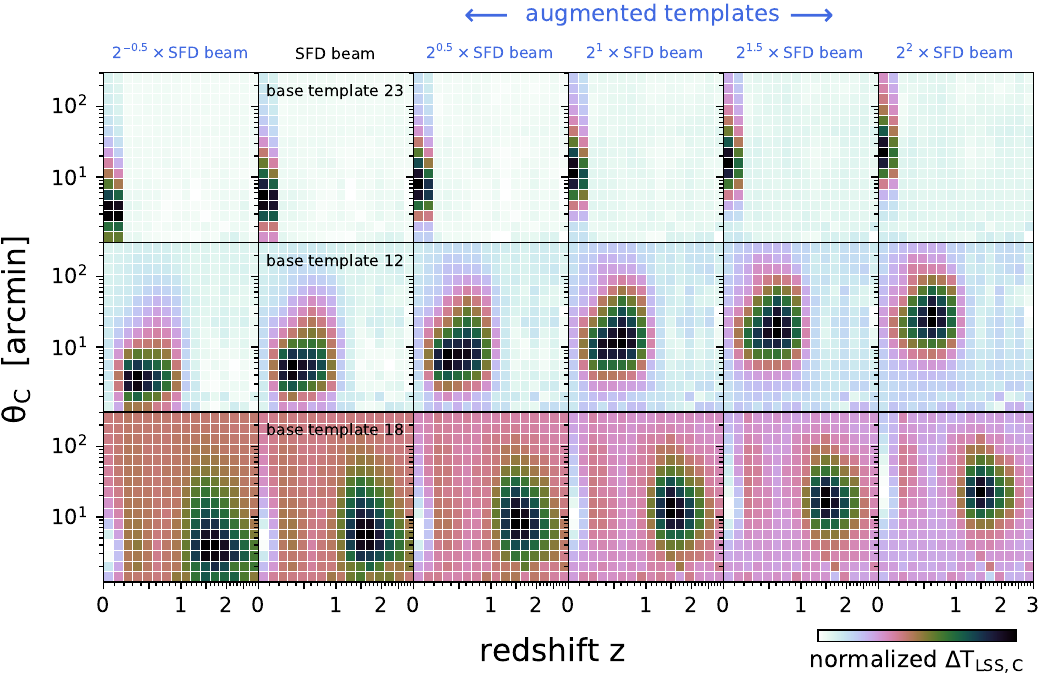}
    \end{center}
    \caption{Same as Figure~\ref{fig:LSSxRef2D} but now for the augmented LSS templates with 6 angular beam sizes for 3 selected base sets (base ID 23, 12, and 18 for the top, middle, and bottom panels, respectively). This plot shows how the beam augmentation shifts angular clustering toward progressively larger scales for larger beams, as expected.}
    \label{fig:LSSxRefBeamAug}
\end{figure*}

\section{Tomographic Statistics in LSS Templates} \label{sec:extract_stats_temp}

In Sections~\ref{sec:extract_stats_SFD_1D} and \ref{sec:extract_stats_SFD_2D}, we extract the 1D plus 2D statistics (Equation~\ref{eq:11} and ~\ref{eq:12}, respectively) of SFD via tomographically cross-correlating the map with SDSS references. Similarly, we need the same sets of statistics for every one of the $180+1$ LSS templates $T_{LSS, i}(\phi)$ built in Appendix~\ref{sec:templates}. This information will then be used to determine the optimal linear coefficients in reconstructing the CIB field in SFD through Equations~\ref{eq:6} to \ref{eq:8}. Differing from measuring the statistics in SFD, here, we do not need to apply further HI-based map-level cleaning because when constructing the LSS templates, stars have been removed. The templates are thus already foreground-free.

Figure~\ref{fig:LSSxRef1D} shows the 1D redshift tomography for the 30 WISE base templates plus one additional point-source template in $\Delta T_{LSS,10'}$ (times a factor $1+z$), the excess dimensionless map values or ``temperature'' of the templates around the SDSS references within $10'$. The middle panel shows the results for those 18 base templates built using CatWISE galaxies split into CMD cells, and the bottom panel shows those 12 built from WISE$\times$SuperCOSMOS sources split in photo-z bins. For CatWISE templates, those using brighter or bluer sources (smaller ID) have their 1D cross-amplitudes peaking at low redshifts, and those from red sources (e.g., ID 10-18) peak at higher redshifts. This is consistent with the expectation that many red sources have their mid-IR fluxes boosted by central AGNs \citep{2005ApJ...631..163S}, and are therefore more likely to still be bright enough to detect at high redshifts. On the other hand, the WISE$\times$SuperCOSMOS templates in the lower panel have more ordered and narrow redshift distributions, as, by construction, they are selected using photo-z information. The beam-augmented templates have redshift distributions $\Delta T_{LSS,10'}(z)$ very similar to their base; we thus do not plot the results here, but they are also used in the reconstruction. As expected, the AKARI FIR point-source template (top panel) is found to be very local at $z<0.1$.

\begin{figure*}[t!]
    \begin{center}
         \includegraphics[width=0.623\textwidth]{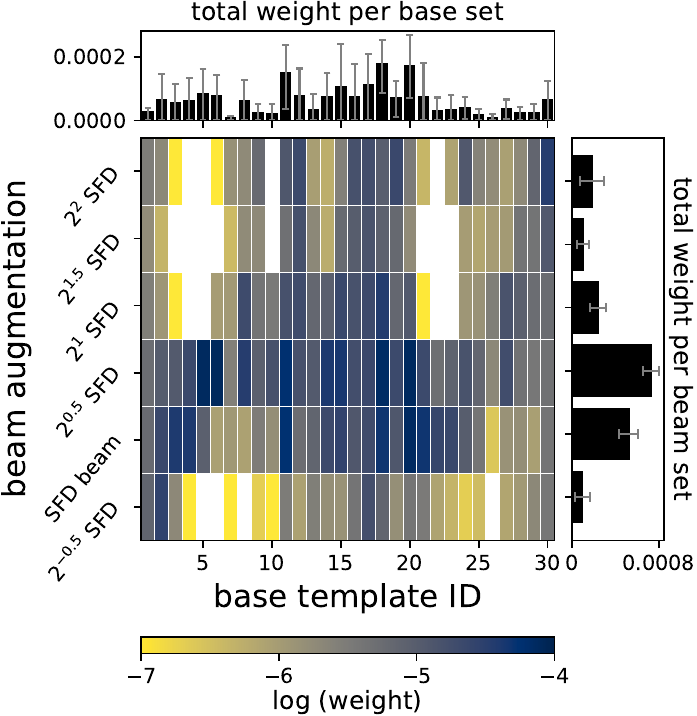}
    \end{center}
    \caption{The best-fit weights, or the linear coefficients $C_i$ for each of the 180 LSS templates (30 bases along the $x$-axis times six augmented beam sizes along the $y$-axis). These weights are obtained by requiring the linearly combined map to reproduce all of the detectable tomographic statistics of the CIB in SFD. The top panel sums over the weights for each base set, showing that all of our WISE galaxy samples contribute. The side panel on the right shows the total weight per beam, indicating that those with a beam slightly larger (factor of $\sqrt{2}$; third row from the bottom) than SFD contribute the most. The errors in the top and side panels are from end-to-end bootstrapping.}
    \label{fig:Weights}
\end{figure*}

\begin{figure}[t!]
    \begin{center}
         \includegraphics[width=0.47\textwidth]{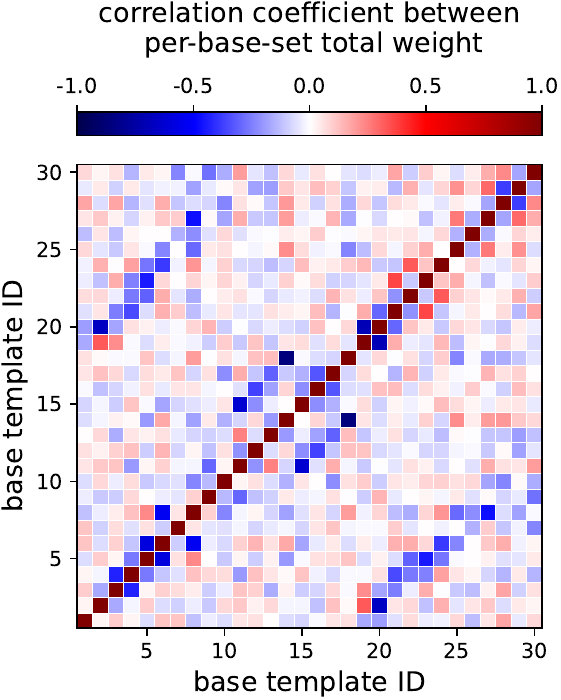}
    \end{center}
    \caption{Correlation coefficient matrix for the weights $C_i$ of the 30 base templates. Anticorrelations (blue) appear when two templates overlap significantly in redshift. Therefore, when one such template is picked in a bootstrap realization, the other one is not needed (e.g., templates ID 2 and 20; see Figure~\ref{fig:LSSxRef1D} for their redshift distributions). These covariances are fully propagated into the final reconstructed LSS/CIB and the associated error maps.}
    \label{fig:Covariance}
\end{figure}

Interestingly, we also see a subtle feature that the reddest CatWISE templates 15--18 appear to be anticorrelated with the SDSS references at $z\lesssim0.1$. This is more likely not physical but a selection effect that a faint, high-redshift object blended with a foreground galaxy would not be detected and reported in the catalog. For our CIB reconstruction, such information would be automatically incorporated the way it should be: since the red templates are slightly biased against, i.e., having negative density contract at the locations of the cosmic web at $z\lesssim0.1$, the blue, low-redshift templates need to compensate by receiving higher weights.

Figure~\ref{fig:LSSxRef2D} shows the 2D tomography $\Delta T_{LSS,C}$ for the 30 base LSS templates using the estimator with compensated filter (Equation~\ref{eq:12}), the same as that applied to SFD in Section~\ref{sec:extract_stats_SFD_2D}. Each panel in Figure~\ref{fig:LSSxRef2D}  is analogous to the 2D plots in the lower B/C panels in Figure~\ref{fig:Summary} for SFD/CSFD. The color scale is linear and, for clarity, normalized to the minimum-to-maximum range of $\Delta T_{C}$ for each template separately. In this plot, we detect both the angular and redshift dependences of the clustering amplitudes, which encode a combination of information for the halo occupation and large-scale linear bias of the galaxies used in each template. In Figure~\ref{fig:LSSxRefBeamAug}, we further show the effect of the beam augmentation using three base sets as examples (ID 23, 12, and 18 in the top, middle, and bottom rows, respectively). From the left to right columns, the beam of these augmented templates increases. We see that the clustering amplitudes are also shifted to large angles accordingly, which is expected by construction.

To this end, the 1D plus 2D cross-statistics for all LSS templates are obtained. This provides the necessary information for the CIB reconstruction in Section~\ref{sec:reconstruction}.

\begin{figure*}[b!]
    \begin{center}
         \includegraphics[width=0.98\textwidth]{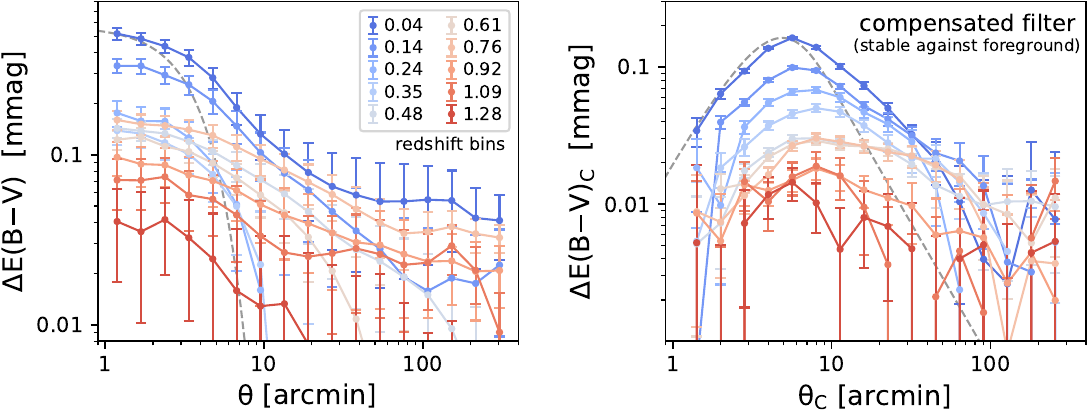}
    \end{center}
    \caption{Comparison between two estimators of angular two-point cross-correlation functions between SFD and the SDSS references in several redshift bins. The left panel uses the usual annulus binning (Equation~\ref{eq:5}), while the right uses the differential compensated filter (Equation~\ref{eq:12}). The gray dashed line in each panel shows the response to a pure Gaussian kernel of the $6.1'$ SFD beam. This figure shows that the compensated filter (right) leads to cross-amplitudes with a smoother progression from low to high redshifts (no blue lines in the middle of reds) as it is more stable against fluctuating local zero-points due to the strong foreground.}
    \label{fig:compensated}
\end{figure*}

\section{Best-fit Coefficients for Reconstruction} \label{sec:best-fit-coef}

Our reconstructed CIB map is a linear combination of the LSS templates (Equation~\ref{eq:6}) that best reproduces the 1D and 2D tomographic statistics of the CIB in SFD, and the fitting is done using the procedure described in Section~\ref{sec:reconstruction}. To gain further insights into the quality of the fit, here we examine the best-fit linear weights $C_i$, an intermediate product of our analysis. Figure~\ref{fig:Weights} shows the posterior mean of these coefficients $C_i$ for the 180 WISE templates organized in the 30 base IDs along the $x$-axis times the six augmented beam sizes along the $y$-axis. The top and side panels show the marginalized weights by summing over the beam ($y$) and base ($x$) axes, respectively, and the error bars are evaluated using our bootstrap realizations of these weights (i.e., standard deviations of ${C_i}^m$ per given $i$). During the fitting, we require a nonnegative coefficient sum for the beam-augmented templates in each base, as the WISE galaxies, no matter which subsample, would not have negative FIR fluxes. Although there is a more than 3-orders-of-magnitude dynamic range in $C_i$, i.e., some templates contribute much more substantially than others, the variances (error bars in the upper band) are actually also large. This is a sign that these templates are highly linearly dependent and overcomplete; although the deprojection to the basis set has no unique solution, such redundancy should allow the combined, reconstructed CIB map to be stable and have small residuals compared to the truth. 

The right-hand side panel shows that the total per-beam weights die off toward both the small and large beam-smearing scales. This suggests that the angular clustering of the CIB should be sufficiently reproduced, and the result is converged, given our template set. Interestingly, the highest weights occur when the templates are smoothed to a slightly more extended, $\sqrt{2}$ times the SFD beam. This confirms the necessity of our beam augmentation and suggests that the CIB has a weaker ``one-halo'' term or a stronger ``two-halo'' term clustering compared to the WISE galaxies used in the templates.

Figure~\ref{fig:Covariance} shows the correlation coefficient matrix for the per-base set weights (same quantity as shown in the upper band in Figure~\ref{fig:Weights}), as evaluated via the spatial block-bootstrapping. Interestingly, we recognize that anticorrelations (blue) occur whenever the templates overlap in redshift. The first place where this is prominent is along the immediate sub- and superdiagonal elements, as templates with adjacent IDs are selected using galaxies with similar colors or photo-z's (see Figure~\ref{fig:LSSxRef1D} for their redshift response). We also notice that many cells with ID 10--18 are anti-correlated since there is more overlap between these high-redshift templates with more extended $P(z)$. Finally, cells in the pair of sidebands around $y = x \pm 18$ are anticorrelated, where 18 corresponds to the number of CatWISE base templates, after which we switch to the WISE$\times$SuperCOSMOS ones starting again from low redshift. This plot demonstrates, again, that the template set has sufficient redundancy for a flexible, empirical reconstruction of the CIB in SFD.

\section{Cross-correlation Optimization}

Due to the strong Galactic foreground in SFD, detecting the LSS imprints with high SNRs requires aggressive optimizations, which we describe here.

\subsection{Compensated Filter}
\label{apd:comp}

First, we describe the need to use the compensated filter (Equation~\ref{eq:12}) in Section~\ref{sec:extract_stats_SFD_2D} when extracting the 2D angular-redshift tomographic statistics as opposed to the classic, plain angular correlation function (Equation~\ref{eq:5}). It is beneficial to consider the picture that the $E(B-V)$ values in SFD around a reference galaxy are the sum of fluctuations of all modes, with the foreground Galactic dust dominating the large-scale modes and the CIB contribution rising toward small scales. The usual angular correlation function in Equation~\ref{eq:5} measures the cumulative clustering (small- plus large-scale modes) at angular separation $\theta$; on the contrary, the compensated filter is a differential estimator that responds only to the small-scale excess clustering within $\theta_C$ compared to the local zero-point between $\theta_C$ and $\sqrt{2}\theta_C$. The compensated filter is thus less affected by mode-mixing, which is advantageous to avoid large-scale biases from the foreground; meanwhile, its covariance matrix in bins of $\theta_C$ would be closer to diagonal. 

\begin{figure*}[t!]
    \begin{center}
         \includegraphics[width=0.98\textwidth]{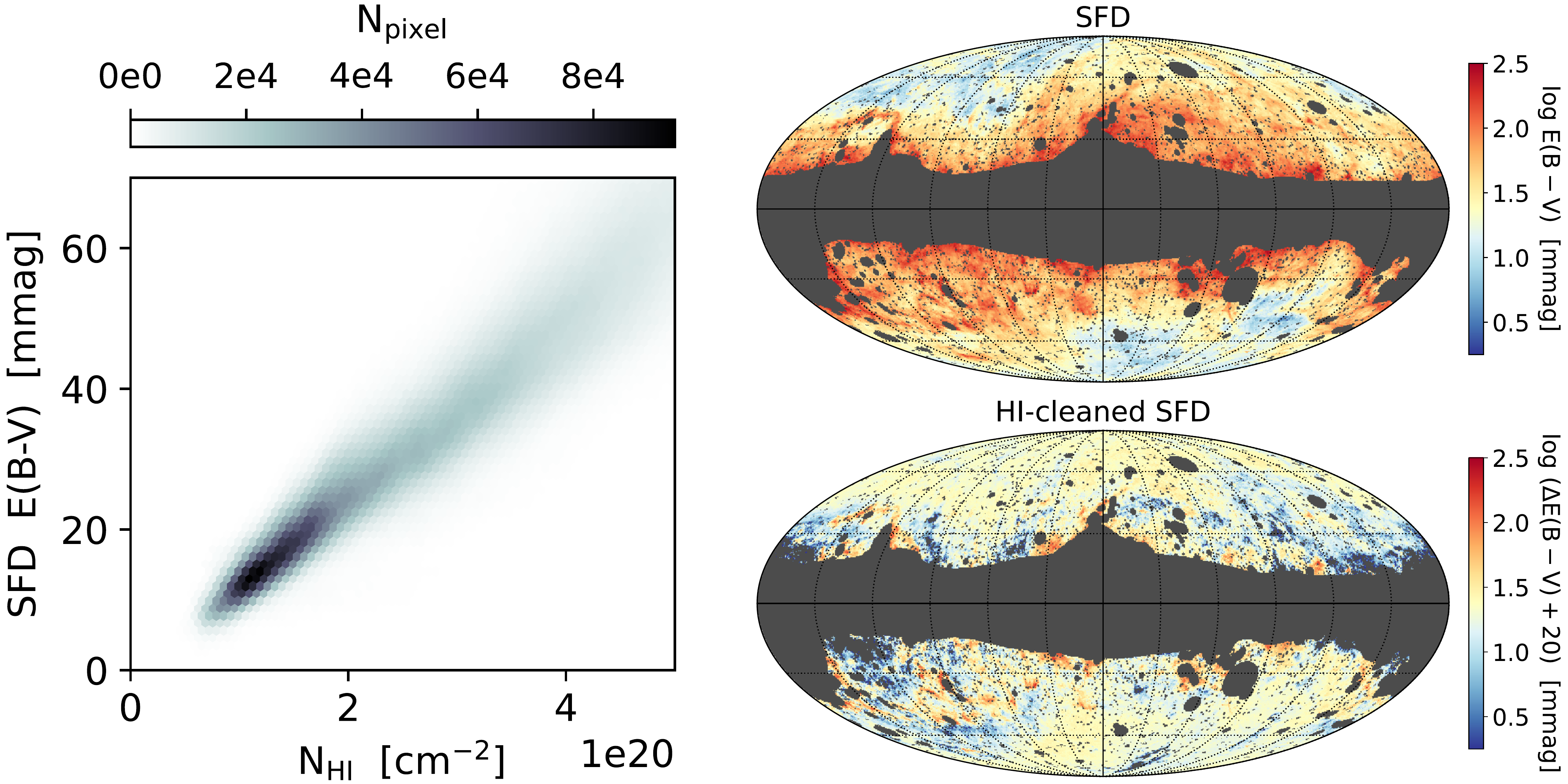}
    \end{center}
    \caption{Pixel value correlation between SFD and the Galactic HI column density map (left panel), the original SFD map (upper right), and the residual SFD map after regressing out the HI-correlated foreground (lower right). As the HI-cleaned SFD, by construction, has a mean close to zero, we add a small offset such that both maps on the right can be plotted over the same logarithmic dynamic range. This plot shows that the HI cleaning  is effective in reducing large-scale foreground gradients.}
    \label{fig:HI_cleaning}
\end{figure*}

\begin{figure*}[t!]
    \begin{center}
         \includegraphics[width=0.62\textwidth]{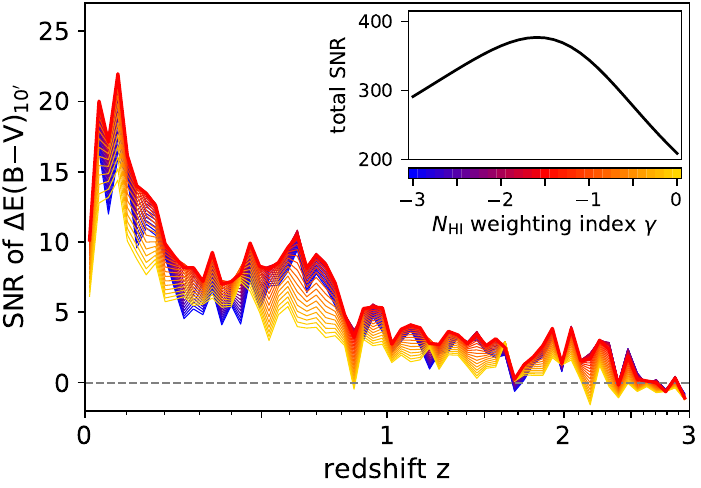}
    \end{center}
    \caption{SNRs for the SFD redshift tomography (cross-correlating with the SDSS references) for different spatial weight maps parameterized by $\gamma$ shown in different color lines. These analyses repeat what is in the upper-B panel in Figure~\ref{fig:Summary} while seeking to optimize the bootstrapping-based SNRs. The main plot shows the per-redshift-bin SNRs and the inset shows the total SNRs summed over redshift for each $\gamma$. Since the noise is related to the foreground, we use Galactic HI to set the family of weight maps $\omega(\phi) \propto N_{\rm HI}^{\gamma}(\phi)$. The case of $\gamma=0$ (yellow line) corresponds to no spatial weighting. An optimal SNR is reached around $\gamma=-1.5$ (thick red line in the main plot), which down-weights regions with a strong foreground. The inset shows that this optimization leads to a factor of 1.8 higher total SNR than that in the no-weighting, $\gamma=0$ case.}
    \label{fig:HI_weighting}
\end{figure*}

In Figure~\ref{fig:compensated}, we compare the two estimators in Equations~\ref{eq:5} and \ref{eq:12} in the left and right panels, respectively; they are applied to measure the excess $E(B-V)$ in SFD around our SDSS references due to the CIB contamination as a function of redshift. Here, data points on the right panel are the same as the color scales of the 2D tomography in the lower panel in Figure~\ref{fig:Summary}-B. Qualitatively, we can see that the two-point functions in the left panel jump quite randomly as the redshift progresses from low to high (blue to red). This is a sign that the large-scale zero-point set by the foreground is unstable, and a slight change in how the SDSS references in each redshift bin sample the sky leads to different mean $E(B-V)$. In contrast, the compensated filter results in the right panel show a smooth progression that the low-redshift measurements (blue) have higher amplitudes than the high-redshift ones (red). Since the covariance of these data points with the compensated filter would be smaller, downstream inference can proceed without a large number of bootstrap sampling. For these reasons, we use the compensated filter for all of our angular tomography.

\subsection{HI-based Foreground Mitigation and Weighting}
\label{apd:foreground}

When extracting the statistics of the extragalactic LSS imprints in SFD/CSFD via tomographic cross-correlations (Figure~\ref{fig:Summary}-B and C) and image stacking (Figures~\ref{fig:stacks-1} and \ref{fig:stacks-DESI}), it is necessary to suppress the foreground for the results to be unbiased and of sufficient SNRs. We adopt a mitigation scheme based on Galactic HI 21~cm emissions. Although HI does not trace dust perfectly, it is a safe foreground template to use with no risk of altering extragalactic signals, as the Galactic 21~cm line is spectrally resolved from the continuum and the redshifted extragalactic HI (see CM19 for a cross-correlation test). We utilize the map of the HI column density within $ |v\rm_{LSR}| < 90\ km\ s^{-1}$ from the HI4PI survey \citep{2016A&A...594A.116H} with $16'$ resolution and consider two aspects of foreground mitigation: (1) map-level cleaning and (2) spatial weighting.

\subsubsection{Map-level Cleaning}
\label{apd:filtering_HI_clean}

We apply the following steps to suppress the foreground in SFD using HI as the template. First, we mask out the noncosmology area as described in Appendix~\ref{apd:base_templates}. Next, we observe a strong linear correlation between the pixel values in the two maps (left panel in Figure~\ref{fig:HI_cleaning}), as both trace the Galactic ISM. We then regress (linearly) out the contribution in the $E(B-V)$ in SFD that can be predicted by the HI column density. Since the HI map is LSS-free, the HI-cleaned SFD should retain all of the LSS imprints. We show the full-sky maps of the original and HI-cleaned SFD in the upper- and lower-right panels, respectively, in Figure~\ref{fig:HI_cleaning}. After the HI cleaning, we see that the spatial gradients due to the Galactic foreground in SFD are largely suppressed. This allows us to have a well-behaved large-scale zero-point in all cross-correlations involving SFD in this work, and our results would be stable against the choice of footprint, i.e., the sky fraction used.

To further suppress the foreground in SFD when extracting the $10'$-scale 1D statistics for redshift tomography in Section~\ref{sec:extract_stats_SFD_1D}, we apply additional high-pass filtering with 1$^{\circ}$ FWHM to SFD. This procedure is also matched for all of the LSS templates $T_{LSS, i}(\phi)$, again, in the 1D case only. This is to ensure that the estimators in both terms in Equation~\ref{eq:7} are exactly the same, which is required for a robust LSS reconstruction. The 1$^{\circ}$ filtering is not applied when the maps are used to extract the 2D statistics (Section~\ref{sec:extract_stats_SFD_2D}), as our 2D estimator with compensated filter (Equation~\ref{eq:12}) is already chosen to access the local zero-point using scales immediately outside $\theta_C$, with a net effect similar to adaptive filtering.  

\subsubsection{Optimal Spatial Weighting}
\label{apd:weight_map}

To maximize the SNRs of cross-correlations in this paper, we further investigate an optimal spatial weighting scheme. We can rewrite our generic tomographic cross-correlation in Equation~\ref{eq:5} to be a weighted version:
\begin{eqnarray}
w_{LSS, R}(\theta, z)\  &=&\  \frac{1}{\langle \omega(\phi) \rangle}\langle \omega(\phi)\, \delta_R(\phi, z) \cdot \Delta I(\phi+\theta) \rangle, \nonumber \\
\label{eq:A1}
\end{eqnarray}
where $\omega(\phi)$ is a weight map whose values need to be specified over the entire cosmology area for cross-correlations (therefore differs from the global scale weight $W(\theta)$ in Equations~\ref{eq:11} and \ref{eq:12}). This leads to weighted equivalents for the 1D and 2D estimators in Equations~\ref{eq:11} and \ref{eq:12}, respectively. These weighted estimators are unbiased under arbitrary weight map $\omega(\phi)$, as long as $\omega(\phi)$ does not correlate with the extragalactic LSS. With the right choice of $\omega(\phi)$, the variance of the estimators can be reduced. A reasonable strategy is to set the weight map to be some $\gamma$ power of the foreground traced, again, by the HI column density map: 
\begin{eqnarray}
\omega(\phi)\ \propto \  N_{\rm HI}^{\gamma}(\phi).
\label{eq:A2}
\end{eqnarray}
We then take an empirical approach by iterating over a range of power indices $\gamma \in [-3, 0]$ and running, end-to-end, the full 1D redshift tomography for SFD.  This basically repeats the cross-correlations in the upper-B panel of Figure~\ref{fig:Summary} with different weight maps. In Figure~\ref{fig:HI_weighting}, we evaluate the SNRs (with errors estimated via block-bootstrapping) for different values of $\gamma$ and show the per-redshift-bin and combined SNRs in the main plot and the inset, respectively. Starting from the no-weighting case, i.e., $\gamma=0$, which is foreground limited, we see that the SNR increases with decreasing $\gamma$. The total SNR reaches the maximum when $\gamma\sim-1.5$, after which we enter the photon-noise-limited regime when the effective area becomes too small at $\gamma\ll-1.5$.  Based on this exercise, we set a spatial weight map of $\omega(\phi) \propto N_{\rm HI}^{-1.5}(\phi)$ for all of our cross-correlations in this paper, including both the 1D and 2D estimators, on SFD, the LSS template maps, and CSFD. Through this optimal weighting scheme, we gain a factor of 1.8 higher SNR compared to the no-weighting case (inset in Figure~\ref{fig:HI_weighting}), equivalent to tripling the amount of data available for free.


\bibliography{CSFD}{}
\bibliographystyle{aasjournal}



\end{document}